\pdfoutput=1

\documentclass[final,5p,times,twocolumn]{elsarticle}

\usepackage{amssymb}

\journal{Nuclear Instrumentations and Methods in Physics Research, Section A}

\begin{document}

\begin{frontmatter}



\title{The Si/CdTe semiconductor Compton camera of the ASTRO-H Soft Gamma-ray Detector (SGD)}


\author[JAXA,UT]{Shin Watanabe}
\author[Nagoya]{Hiroyasu Tajima}
\author[Hiroshima]{Yasushi Fukazawa}
\author[JAXA,UT]{Yuto Ichinohe}
\author[JAXA]{Shin'ichiro Takeda}
\author[RIKEN]{Teruaki Enoto}
\author[JAXA,UT]{Taro Fukuyama}
\author[Hiroshima]{Shunya Furui}
\author[MHI]{Kei Genba}
\author[JAXA,UT]{Kouichi Hagino}
\author[JAXA]{Atsushi Harayama}
\author[MHI]{Yoshikatsu Kuroda}
\author[MHI]{Daisuke Matsuura}
\author[Hiroshima]{Ryo Nakamura}
\author[UT]{Kazuhiro Nakazawa}
\author[UT]{Hirofumi Noda}
\author[JAXA]{Hirokazu Odaka}
\author[JAXA]{Masayuki Ohta}
\author[MHI]{Mitsunobu Onishi}
\author[JAXA,UT]{Shinya Saito}
\author[WASEDA,JAXA]{Goro Sato}
\author[JAXA,UT]{Tamotsu Sato}
\author[JAXA,UT]{Tadayuki Takahashi}
\author[Kyoto]{Takaaki Tanaka}
\author[JAXA,UT]{Atsushi Togo}
\author[Nagoya]{Shinji Tomizuka}

\address[JAXA]{Institute of Space and Astronautical Science, Japan Aerospace Exploration Agency, 3-1-1 Yoshinodai, Chuo Sagamihara, Kanagawa, 252-5210, Japan}
\address[UT]{Department of Physics, The University of Tokyo, 7-3-1 Hongo, Bunkyo, Tokyo, 113-0033, Japan}
\address[Nagoya]{Solar-Terrestrial Environment Laboratory, Nagoya University, Furo, Chikusa Nagoya, Aichi, 464-8601, Japan}
\address[Hiroshima]{Department of Physical Science, Hiroshima University, 1-3-1 Kagamiyama, Higashi-Hiroshima, Hiroshima, 739-8526, Japan}
\address[RIKEN]{Nishina Center, RIKEN, 2-1 Hirosawa, Wako, Saitama, 351-0198, Japan}
\address[MHI]{Nagoya Guidance and Propulsion Systems Works, Mitsubishi Heavy Industry Ltd., 1200 Higashi Tanaka, Komaki, Aichi, 485-8561, Japan}
\address[WASEDA]{Research Institute for Science and Engineering, Waseda University, 3-4-1 Okubo, Shinjuku, Tokyo 169-8555, Japan}
\address[Kyoto]{Department of Physics, Kyoto University, Kitashirakawaoiwake, Sakyo Kyoto, Kyoto, 606-8502, Japan}

\begin{abstract}
 The Soft Gamma-ray Detector (SGD) is one of the instrument payloads onboard ASTRO-H, 
and will cover a wide energy band (60--600 keV) at a background level 
10 times better than instruments currently in orbit. 
The SGD achieves low background by combining a Compton camera scheme 
with a narrow field-of-view active shield. 
The Compton camera in the SGD is realized as a hybrid semiconductor detector system 
which consists of silicon and cadmium telluride (CdTe) sensors. 
The design of the SGD Compton camera has been finalized and the final prototype, which has the same 
configuration as the flight model, has been fabricated for performance evaluation. 
The Compton camera has overall dimensions of 12~cm~$\times$~12~cm~$\times$~12~cm, consisting of 32 layers of Si pixel sensors 
and 8 layers of CdTe pixel sensors surrounded by 2 layers of CdTe pixel sensors. 
The detection efficiency of the Compton camera reaches about 15\% and 3\% for 100~keV and 511~keV gamma rays, respectively.
The pixel pitch of the Si and CdTe sensors is 3.2 mm, and the signals from all 13312 pixels are processed 
by 208 ASICs developed for the SGD. 
Good energy resolution is afforded by semiconductor sensors and low noise ASICs, 
and the obtained energy resolutions with the prototype Si and CdTe pixel sensors are 
1.0--2.0 keV (FWHM) at 60 keV and 1.6--2.5 keV (FWHM) at 122 keV, respectively. 
This results in good background rejection capability due to better constraints on Compton kinematics. 
Compton camera energy resolutions achieved with the final prototype are 
6.3~keV (FWHM) at 356~keV and 10.5~keV (FWHM) at 662~keV, respectively, 
which satisfy the instrument requirements for the SGD Compton camera (better than 2\%).
Moreover, a low intrinsic background has been confirmed by the background measurement with the final prototype.

\end{abstract}


\end{frontmatter}


\section{ASTRO-H SGD}
\label{sec:astroh_sgd}
ASTRO-H, the new Japanese X-ray Astronomy Satellite\cite{NeXT08, Takahashi10, Takahashi12} following the currently-operational Suzaku satellite, aims to fulfill the following scientific goals:
\begin{itemize}
\item Revealing the large-scale structure of the universe and its evolution.
\item Understanding the extreme conditions of the universe.
\item Exploring the diverse phenomena of the non-thermal universe.
\item Elucidating dark matter and dark energy.
\end{itemize}
In order to fulfill the above objectives, the ASTRO-H satellite hosts the following four types of instruments: 
SXT (Soft X-ray Telescope) + SXS (Soft X-ray Spectrometer), SXT + SXI (Soft X-ray Imager), 
HXT (Hard X-ray Telescope) + HXI (Hard X-ray Imager)\cite{Takahashi04NAR,Takahashi02-NeXT,Takahashi02,Takahashi04-SGD,HXI08,Kokubun10,Kokubun10spie,Kokubun12spie}  
and SGD (Soft Gamma-ray Detector) \cite{Takahashi04NAR,Takahashi02-NeXT,Takahashi02,Takahashi04-SGD,Kokubun10,Tajima05,Tajima10,Watanabe12spie}.

The SGD will cover the energy range of 60--600~keV with a high sensitivity.
The SGD utilizes semiconductor detectors using Si and CdTe pixel 
sensors with good energy resolution ($\lesssim$2~keV) for the Compton camera, 
which were made possible by recent progress on the development of 
low noise Si gamma-ray sensors\cite{Tajima02,Tajima03,Fukazawa05_si,Takeda07} and
high quality CdTe sensors\cite{Takahashi01b,Watanabe05,Watanabe09,Takeda09,Takeda12,CdTebook1,CdTebook2}.
The BGO active shield provides a low background environment by rejecting the
majority of external backgrounds.  Internal backgrounds are rejected based 
on the inconsistency between the constraint on the incident angle of 
gamma rays from Compton kinematics and that from the narrow FOV (field 
of view) of the collimator.  This additional background rejection by 
Compton kinematics will improve the sensitivity by an order of magnitude 
in the 60--600~keV band compared with the currently operating space-based instruments.

In this paper, we will present the detailed configuration and the data acquisition system of the SGD Compton camera. 
We will also present the performance evaluated for the final prototype, which is equivalent to the flight model.

\section{SGD concept and Si/CdTe Semiconductor Compton Camera}
\label{sec:sgd_concept_sicdtecc}

\begin{figure}[bth]
   \centering
   \includegraphics[width=0.45\textwidth]{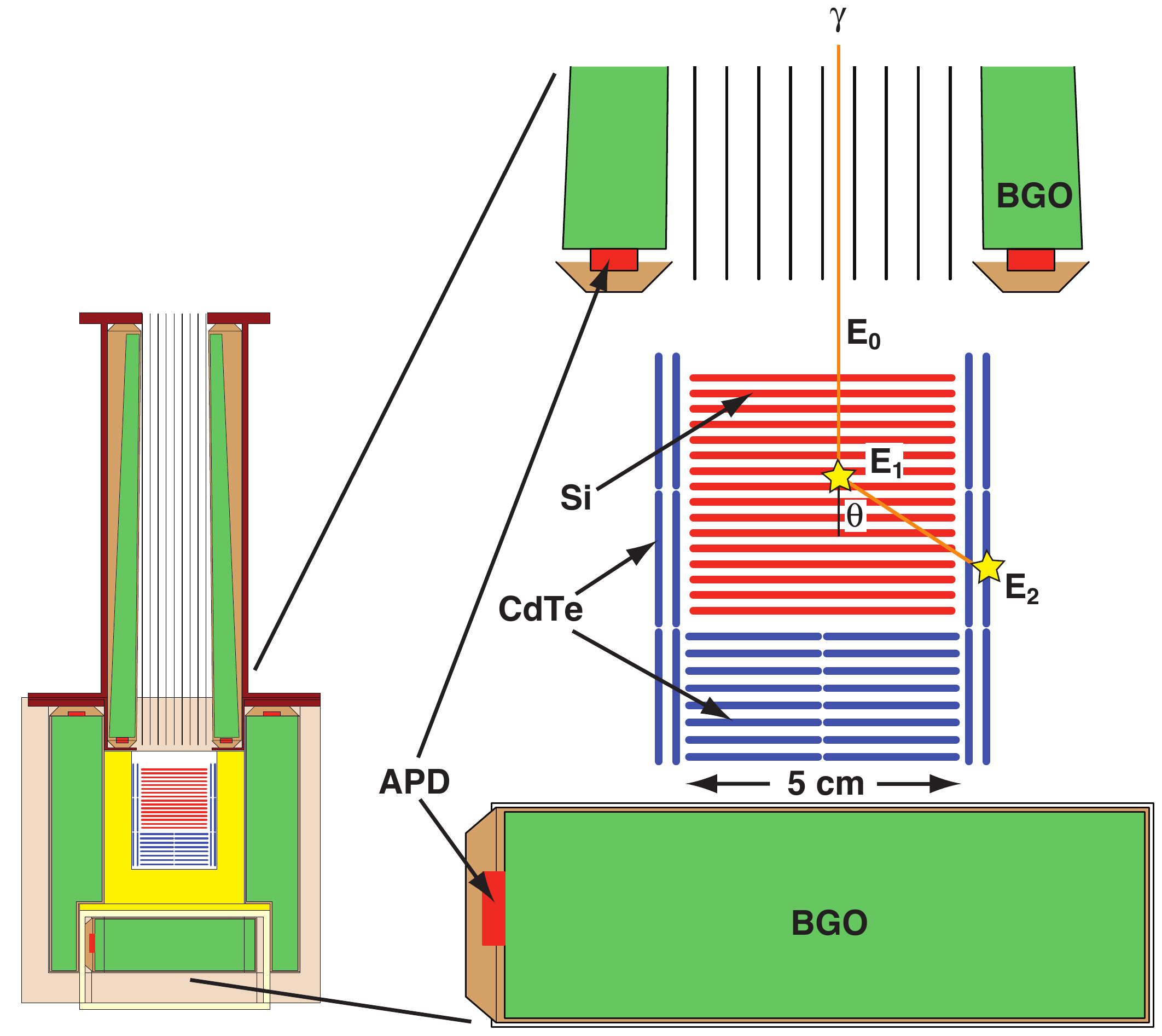}
    \caption{Conceptual drawing of an SGD Compton camera unit.}\label{fig:SGD-concept}
   \end{figure}

The SGD is based on the concept of narrow FOV Compton telescopes\cite{Takahashi02-NeXT}, combining Compton cameras and active
well-type shields.
The active well-type shield concept originates from the Hard X-ray Detector (HXD)\cite{HXD} onboard the Suzaku satellite.
The HXD achieves the best sensitivities in the hard X-ray band, 
consisting of Si photodiodes and GSO scintillators with BGO active shield and copper passive collimator.
The SGD, however, replaces the Si photodiodes and GSO scintillators with the Compton camera, 
which provides additional information for the background rejection.
Figure~\ref{fig:SGD-concept} shows a conceptual drawing of an SGD unit.
A BGO collimator defines a field of view of $\sim$10$^{\circ}$ for high energy photons 
while a fine collimator restricts the FOV to $\lesssim$0.6$^{\circ}$ for low energy photons ($\lesssim$150~keV), 
which is essential to minimize the CXB (cosmic X-ray background) and source confusion.
Scintillation light from the BGO crystals is detected by avalanche photo-diodes (APDs) 
allowing for a compact design compared to phototubes.

The hybrid design of the Compton camera module incorporates both Si and CdTe imaging detectors.
The Si sensors are used as detectors for Compton scattering 
since Compton scattering is the dominant process in Si above $\sim$50~keV compared with $\sim$300~keV for CdTe.
The Si sensors also provide better constraints on the Compton kinematics because of smaller effect of Doppler broadening\cite{Ribberfors19752067,Zoglauer20021302}.
The CdTe sensors are used to absorb gamma rays following Compton scattering in the Si sensors.

\section{Instrument-level requirements for SGD Compton cameras}

The ASTRO-H mission-level science objectives described above require the SGD to provide 
spectroscopy up to 600~keV for over 10 accreting supermassive black holes with fluxes equivalent to 
1/1000 of the Crab Nebula (as measured over the 2--10~keV band, assuming the spectrum 
to be a power-law with spectral index of 1.7).  
This mission-level science requirement 
defines the following instrument-level requirements for the SGD Compton camera:  
\begin{itemize}
\item Effective area for the detector must be greater than 20~cm$^2$ at 100~keV to obtain a sufficient number of photons in a reasonable observation time (typically 100~ks).
\item Observation energy range must be from 60~keV to 600~keV.
\item Energy resolution must be better than 2~keV (FWHM) or better than 2\% (FWHM).
\end{itemize}

There are the following design constraints on the SGD Compton camera fro mounting the ASTRO-H:
\begin{itemize}
\item The size of one camera must be 12~cm~$\times$~12~cm~$\times$~12~cm to minimize the size 
of BGO active shield, since the BGO is the dominant contributor to the total weight of the SGD.
\item The number of Compton cameras must be six in total ASTRO-H.
\item Power consumption must be lower than 6~W for one camera.
\end{itemize}

The instrument-level requirements and the design constraints described above guide the designs of the Si sensor,
the CdTe sensor and the readout Application Specific Integrated Circuit (ASIC) for both.
The detection area of the Si sensor must be larger than 5~cm~$\times$~5~cm. 
The total thickness of the Si sensor must be about 2~cm, which corresponds to the 50\% interaction efficiency for 100~keV photons. Therefore, 32 layers of Si sensors are needed when 0.6~mm thick Si devices are used.
In order to satisfy the effective area requirement, CdTe sensors must cover 50\% of the solid angle covered by the Si sensors. 
The readout ASICs for the Si and CdTe sensors must have an internal analog-to-digital converter (ADC) and 
must be controllable with digital signals
because space in the Compton camera is limited. Moreover, the ASIC must consume less than 0.5~mW/channel
and have good noise performance of 100--200~e- (ENC) under the condition that the input capacitance is several~pF.

\section{SGD Compton Camera Design}
\label{sec:cc_design}

\subsection{Overall Design}
\label{subsec:cc_overall_design}
\begin{figure}[htbp] 
   \centering
   \begin{tabular}{c}
   (a) \\
   \includegraphics[width=0.425\textwidth]{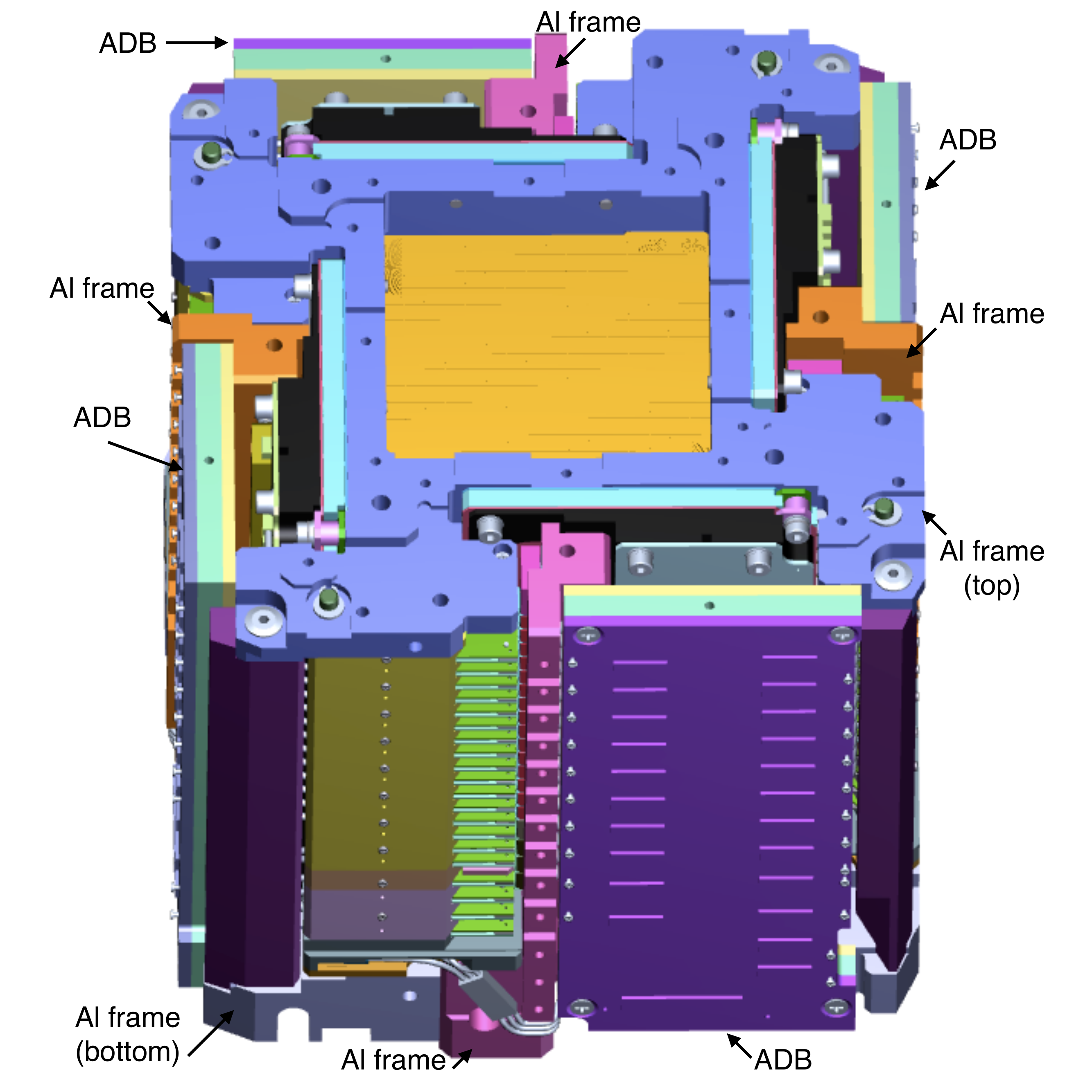} \\
      \\
   (b) \\
      \includegraphics[width=0.425\textwidth]{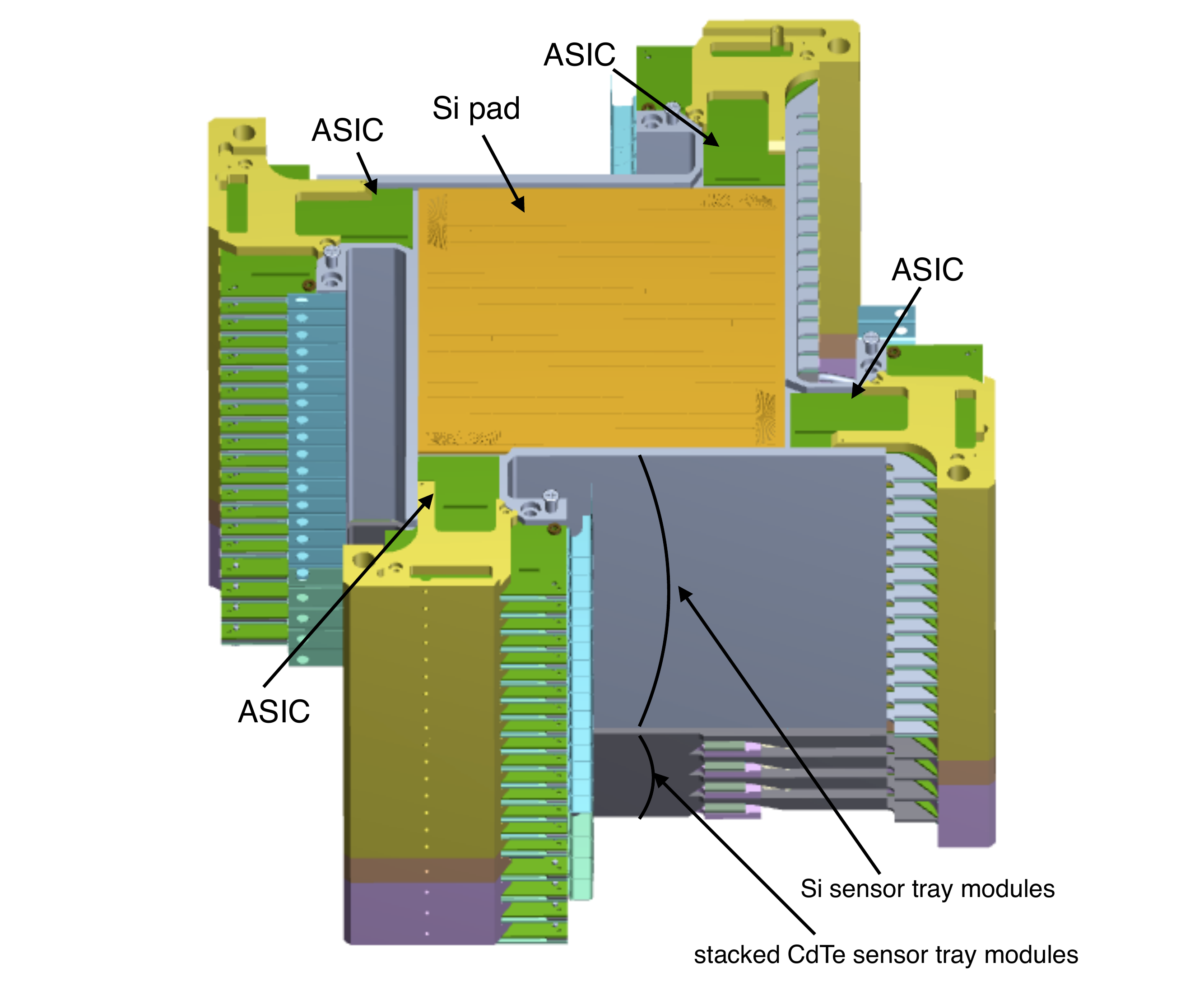} \\
      \\
   (c) \\
   \includegraphics[width=0.425\textwidth]{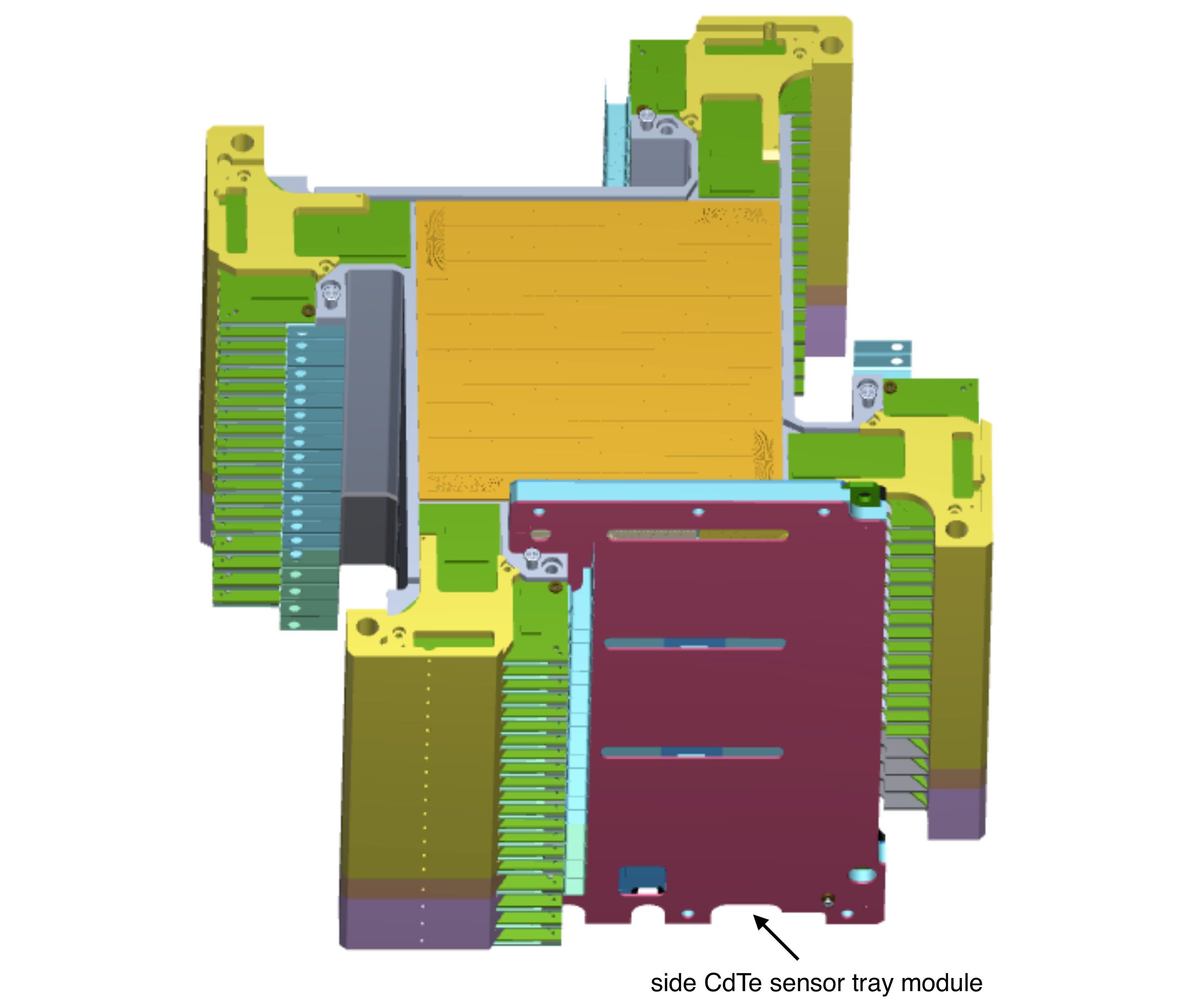} 
   \end{tabular}
   \caption{(a) 3D model of Compton camera structure. (b) Si and CdTe stack only. 
   (c) One side CdTe tray module is installed in the stack.}
   \label{fig:CC-structure}
\end{figure}

Based on the design guide described in the last section,
the Compton camera consists of 32 layers of Si sensors 
and 8 layers of CdTe sensors surrounded by 2 layers of 
CdTe sensors. Figure~\ref{fig:CC-structure} shows a 
3D model of the Compton camera structure.
This arrangement allows a placement of the CdTe sensor 
on the side very close to the stacked Si and CdTe sensors, 
maximizing the coverage of the photons scattered by the Si sensors.
In addition to sensor modules, the Compton camera holds 
an ASIC controller board (ACB) and four ASIC driver boards (ADBs).
The ACB holds a field programmable gate array (FPGA) 
that controls the ASICs. The ADB buffers control signals 
from the ACB, sends control signals to 52 ASICs, and also 
provides a current limiter to power the ASICs.

The mechanical structure of the Compton camera needs to hold all components described above within a volume of $12\times12\times12$~cm$^3$. 
Another important requirement 
for the mechanical structure is sensor cooling.
All sensors need to have a temperature that is within 5$^{\circ}$C of 
the cold plate interface at the bottom of the Compton camera.

Figure~\ref{fig:CC-structure} (a) shows the mechanical support structure of the Compton camera.
The Compton Camera consists of a stack of Si and CdTe sensor trays as shown 
in Figure~\ref{fig:CC-structure} (b). In addition to the stack, one "side CdTe sensor module" is installed on each side
as shown in Figure~\ref{fig:CC-structure} (c). 
Each ADB is attached to the side CdTe sensor module and an ACB is attached to the bottom frame.
The material of the sensor tray structure employs polymide. 

\subsection{Readout ASIC for Si and CdTe sensors}
\label{subsec:asic}
The main performance requirements for an ASIC are low noise and low power. 
In order to satisfy these main requirements, the ASIC was developed based on the VIKING 
architecture\cite{VA94,Tajima04} which is known for good noise performance 
and has been in various space experiments like Swift\cite{Barthelmy2005143}, PAMELA\cite{Picozza2007296} and AGILE\cite{Tavani2009995}.

\begin{figure*}[hbtp] 
   \centering
   \includegraphics[width=0.825\textwidth]{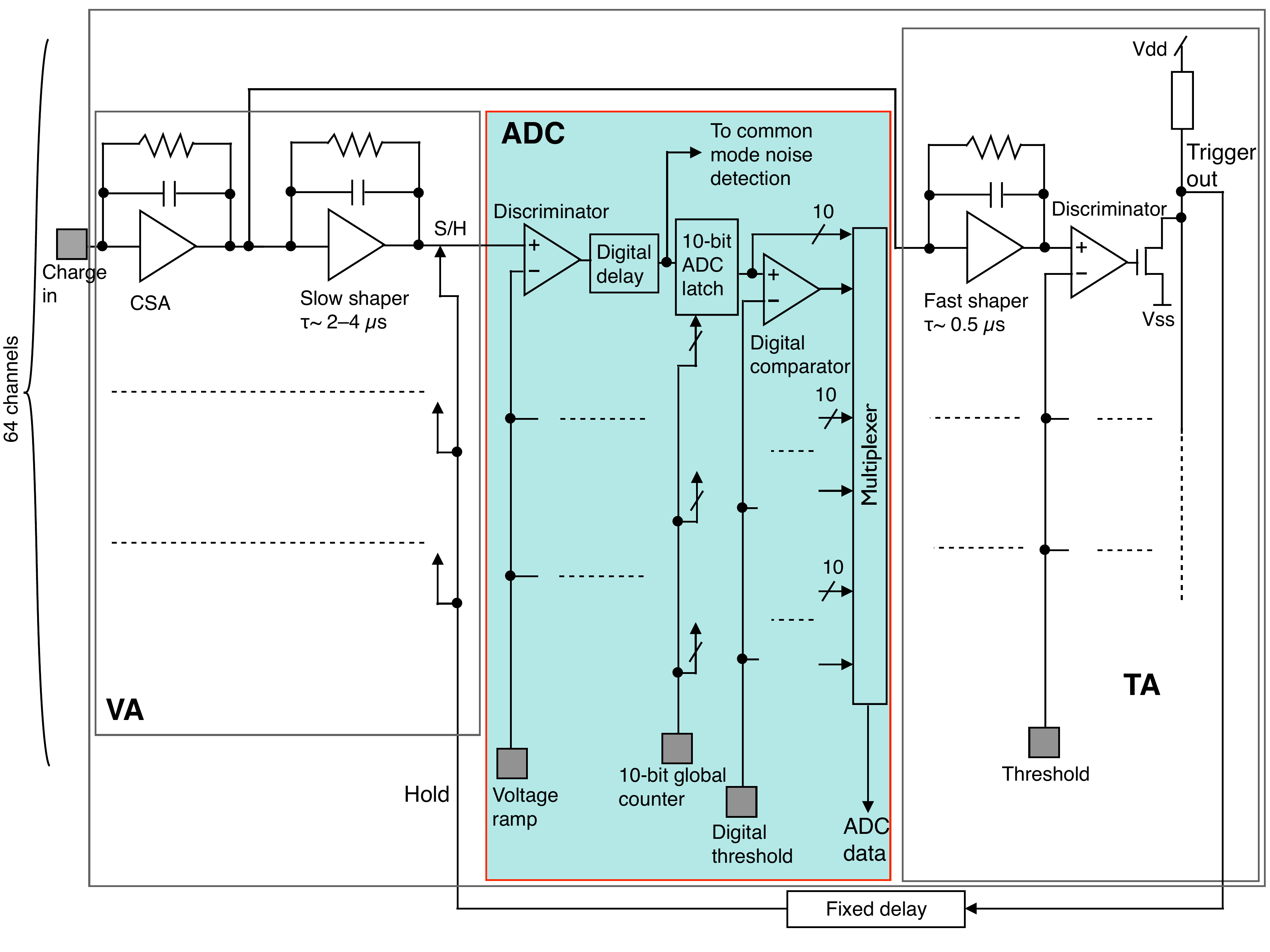} 
   \caption{Circuit diagram of the ASIC developed for the SGD. The circuits shown in a blue background are implemented in this development.}
   \label{fig:VIKING}
\end{figure*}

Figure~\ref{fig:VIKING} shows the circuit diagram for the ASIC developed for the SGD (and HXI) of ASTRO-H.
Each channel consists of charge sensitive amplifier followed by two shapers. 
One shaper with a short shaping time is followed by a discriminator 
to form a trigger signal.
The other shaper with a long shaping time is followed by a sample and 
hold circuit to hold the pulse height at the timing specified by an external hold signal.

Many important functionalities are integrated in the ASIC for the SGD in order to minimize additional components required to read out the signal, as shown in the circuit diagram with a blue background region.
The signals in all channels are converted to 
digital values concurrently with Wilkinson-type analog-to-digital 
converters (ADCs). The conversion time is less than 100~$\mu$s.
In order to minimize the readout time, the only channels that are read out 
are those above a data threshold that can be digitally set for each 
channel independently from the trigger threshold.
We usually observe common mode noise from this type of ASIC at the level 
of $\sim$1~keV.
Common mode noise has to be subtracted to accurately 
apply the threshold for the zero suppression.
The common mode noise level of each event is detected by taking an 
ADC value of the 32nd [a half of total number of channels] pulse height, 
corresponding to a median value of all ADC values.
With zero suppression, the readout time is $0.5\;\mu$s per ASIC when 
no data is readout and $(9+n)\;\mu$s when we readout $n$ channels. 
Without zero suppression, the readout time becomes $73\;\mu$s per ASIC. 

The ASIC produces all necessary analog bias currents and voltages 
on the chip by internal Digital to Analog Converters (DACs) except 
for the main bias current which sets the scale of all bias currents. 
Each bit of the registers for all internal DACs and other functions 
consists of three flip-flops and a majority selector for tolerance 
against single event upset (SEU).
If the majority selector detects any discrepancies among three 
flip-flops, it will set a SEU flag which will be readout as a 
part of output data.
The ASIC is fabricated on a wafer with an epitaxial layer that 
will improve immunity against latch up.
Table~\ref{table:ASIC-spec} summarizes the specifications. 

\begin{table}[htdp]
\caption{SGD ASIC specifications}
\begin{center}
\begin{tabular}{|l|r|}
\hline
\multicolumn{2}{|c|}{Geometrical specifications} \\\hline
Number of channels & 64 \\
Input pitch & 91~$\mu$m \\
Thickness & 0.45~mm \\ \hline
\multicolumn{2}{|c|}{Analog specifications} \\\hline
Power consumption & 0.2~mW/channel \\
Fast shaper peaking time & 0.6~$\mu$s \\
Slow shaper peaking time & $\sim$3 $\mu$s \\
Noise performance & 180~$e^-$ (RMS) at 6~pF load\\
& 1.5~keV (FWHM) for Si \\
Threshold & 1500~$e^-$ at 6~pF load \\
& 5.4 keV for Si \\
Threshold range & 625 -- 6250~$e^-$ \\
Threshold step & 208~$e^-$ \\
Dynamic range & $\pm$100,000~$e^-$ \\
& 360~keV for Si \\ \hline
\multicolumn{2}{|c|}{Digital specifications} \\\hline
ADC setup time & 5~$\mu$s \\
ADC power consumption & 0.5--2 mW/channel \\
& 5--20~$\mu$W/channel at 100~Hz \\
Data clock speed & $<$10~MHz \\
Conversion clock speed & $<$10~MHz (external clock) \\
& $<$20~MHz (internal clock)\\
Conversion time & $<$100~$\mu$s (external clock) \\
& $<$50~$\mu$s (internal clock)\\
Readout time (no data) & 0.5~$\mu$s per ASIC \\
Readout time ($n$ channels) & $(9+n)$~$\mu$s per ASIC \\ \hline
\end{tabular}
\end{center}
\label{table:ASIC-spec}
\end{table}%

The data input and output circuits on the ASIC are designed to allow 
daisy-chaining of multiple ASICs.  In one scheme, the data output 
of one ASIC can be connected to the input of another ASIC and the 
ASIC will pass the input data to the output via a shift register.
This scheme is used to set register values.
In another scheme, the outputs of several ASICs can be connected to a single bus.
The output is controlled by passing a token from ASIC to ASIC.
Or, in the case of trigger signal, ASICs can issue trigger signals at any time since the output circuit is open-drain FET to allow multiple triggers on the same bus.
In the Compton camera, 6 ASICs (in the side CdTe tray modules) or 8 ASICs (in the stack tray modules) are daisy chained.

\subsection{Si and CdTe Pad Detector}
\label{subsec:sicdtepad}

Si and CdTe sensors are pixellated to give two-dimensional coordinates 
with a pixel size of $3.2\times3.2$~mm$^{2}$. Pixel size is chosen so as 
to minimize the number of pixels for lower 
power consumption while avoiding the pixel size to be the dominant 
contribution to the angular resolution of Compton kinematics.

\begin{figure*}[htbp] 
   \centering
   \begin{tabular}{ll}
   (a) & (b)\\
   \includegraphics[width=0.3\textwidth]{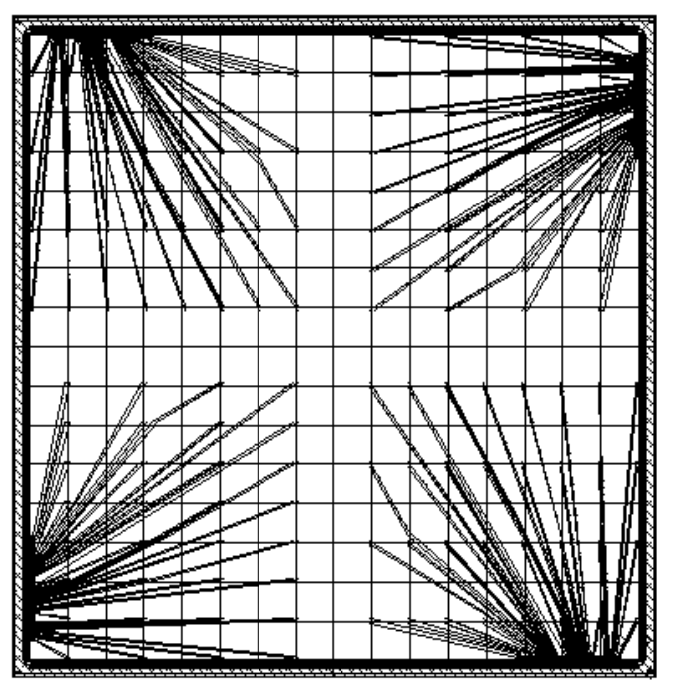} &
   \includegraphics[width=0.45\textwidth]{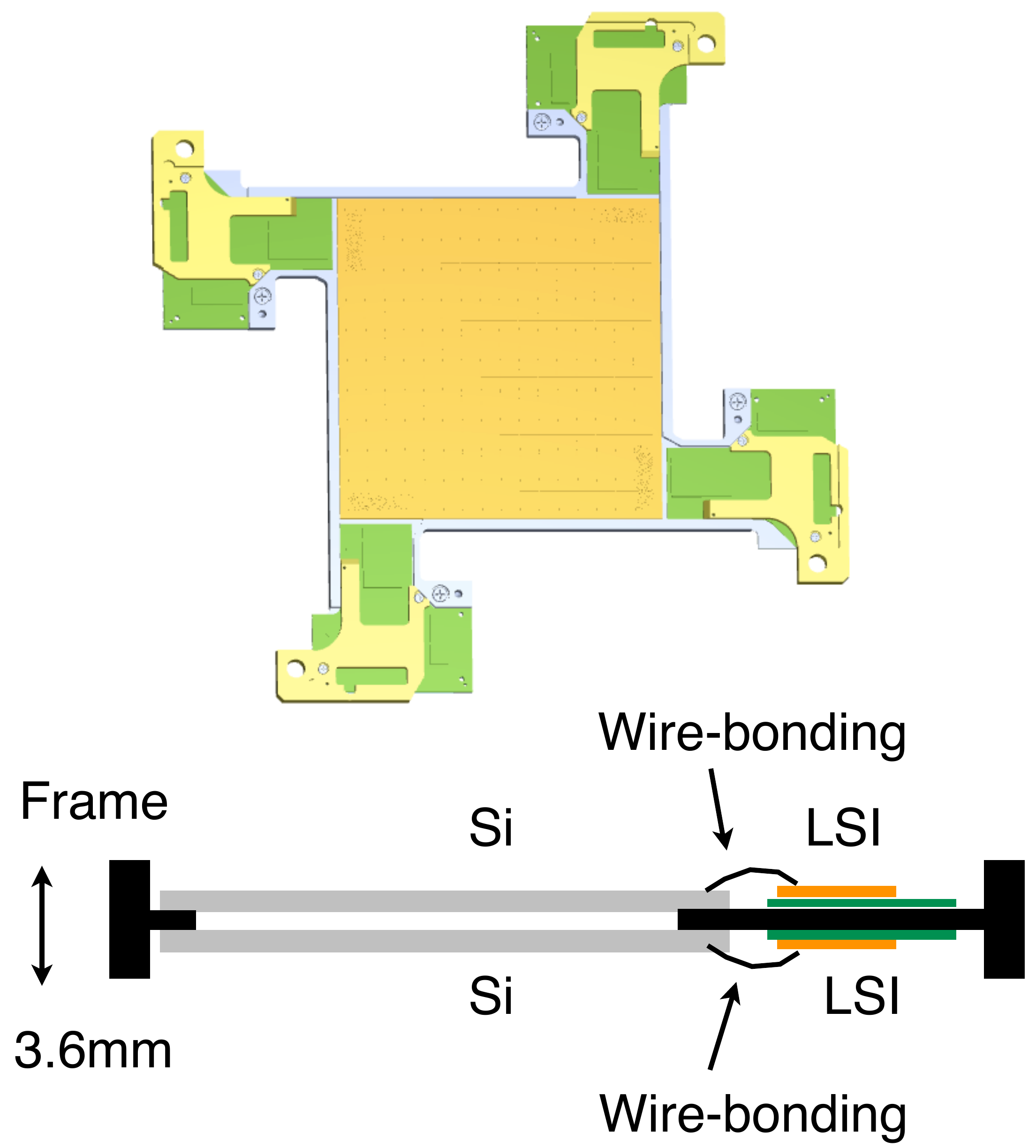}\\
   \end{tabular}   
   \caption{(a) Schematic drawing of Si sensor showing layout of pixels and readout traces. (b) 3D model and conceptual illustration of the Si sensor tray module.}
   \label{fig:si_sensors}
\end{figure*}

The Si pad devices are manufactured by Hamamatsu Photonics K.~K.
Each Si sensor has $16\times16$ pixels providing $5.12\times5.12$~$\mathrm{cm}^2$ active area.
A signal from each pixel on the Si sensor is brought out to one of the bonding 
pads at the corner of the sensor by a readout electrode placed 
on top of the SiO$_2$ insulation layer with a thickness of 1.5~$\mu$m 
as shown in Figure~\ref{fig:si_sensors} (a).
The readout electronics are DC-coupled to the Si sensor.
The thickness of each Si device is 0.6~mm and the operating bias voltage is 230~V. 
The capacitance per pad is 4--15~pF, including the capacitance attributed to the traces.
The leakage current is about 2~nA for one device at the temperature of $-$20$^{\circ}$C .

\begin{figure*}[htbp] 
   \centering
   \begin{tabular}{ll}
   (a) & (b)\\
   \includegraphics[width=0.48\textwidth]{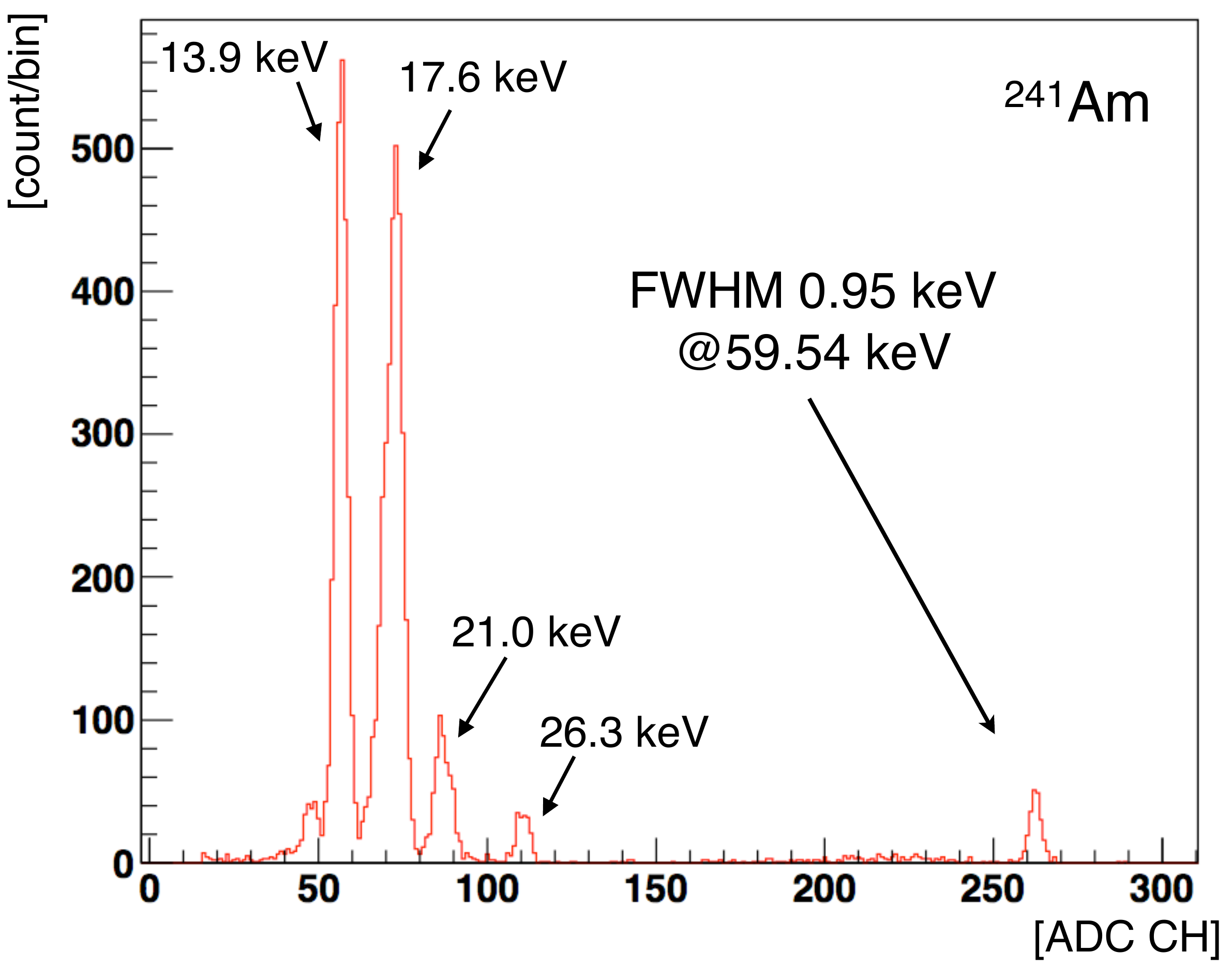} &
   \includegraphics[width=0.45\textwidth]{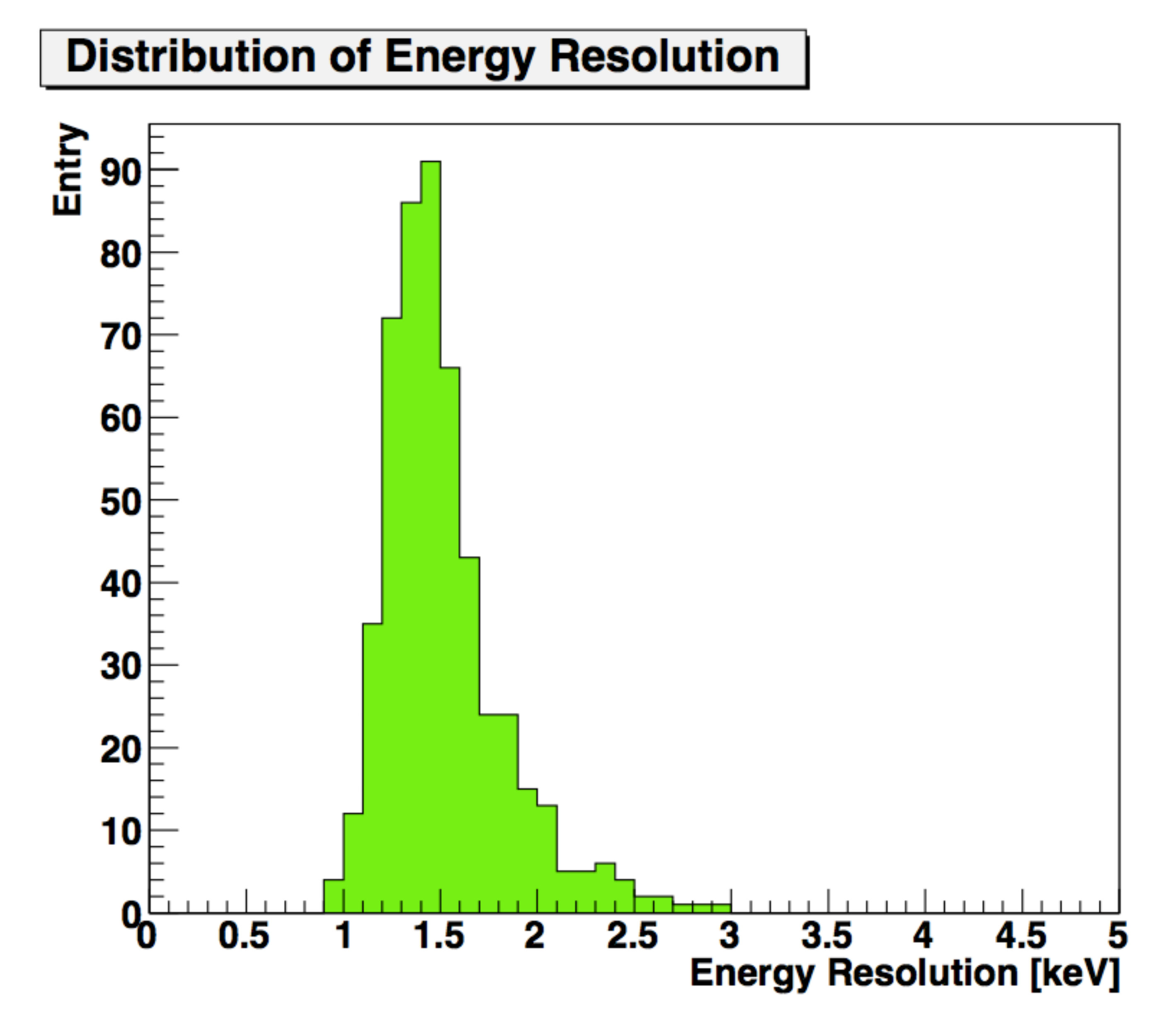} 
   \end{tabular}   
   \caption{(a) $^{241}$Am X-ray/gamma-ray spectrum obtained with a Si sensor of a Si tray module. This spectrum is obtained with one of the best pixels. The FWHM energy resolution is 0.95~keV for 59.54~keV gamma rays. The operating temperature is $-$20$^{o}$C. (b) The distribution of the energy resolutions obtained with the pixels of the Si sensor tray modules for 59.54~keV gamma rays. The mean energy resolution is 1.5~keV (FWHM).}
   \label{fig:sisensors_performance}
\end{figure*}

Figure~\ref{fig:si_sensors} (b) shows the structure of the Si sensor tray module. 
One tray module consists of two Si pad devices and eight front end cards (FECs) with one ASIC.
Sixteen tray modules are stacked in one Compton camera.
The spectral performance obtained with Si sensor tray modules is shown in Figure~\ref{fig:sisensors_performance}.
Figure~\ref{fig:sisensors_performance} (a) shows the $^{241}$Am spectrum obtained with one of the best pixels. 
The FWHM energy resolution is 0.95~keV for 59.54~keV gamma rays at the operating temperature of $-$20$^{o}$C.
Figure~\ref{fig:sisensors_performance} (b) shows the distribution of the energy resolutions obtained 
with the pixels of the Si sensors. This variation of the energy resolutions result from the difference of 
input capacitance due to readout traces for readout signal from pixels.

The CdTe pad devices are produced and processed by ACRORAD Co. Ltd.
The CdTe sensor has $8\times8$ pixels providing $2.56\times2.56$~$\mathrm{cm}^2$  
active area as it is difficult to fabricate a CdTe sensor much 
larger than $3\times3$~$\mathrm{cm}^2$.  CdTe sensors are tiled in a $2\times2$ 
array for each layer in the bottom and in a $2\times3$ array for each 
layer on the side to obtain the required active area.
In order to overcome small mobility and short lifetime of carriers 
in CdTe sensors, we employ a Schottky-barrier diode type CdTe sensor 
with Indium (In) anode and Platinum (Pt) cathode so that we can apply high bias voltage with low leakage current.
The indium electrode functions as a common biasing electrode while Pt 
electrodes form pixels. 
Titanium is placed on the In electrode to reduce the resistance.
Diode type CdTe sensors suffer degradation of energy resolution due to charge trapping 
over time (i.e. polarization). 
It is known that the polarization slows down at lower temperature and the effect of 
polarization can be reduced by applying higher bias voltage.
For example, it was found that one week of operation of this type of CdTe 
sensor shows little polarization effect at $<$5$^{\circ}$C 
and $>$1000~V/mm.
Moreover, this polarization effect can be recovered 
by turning off the bias voltage and 
the recovery process accelerates at a higher temperature.

\begin{figure}[hbtp] 
   \centering
   \includegraphics[width=0.48\textwidth]{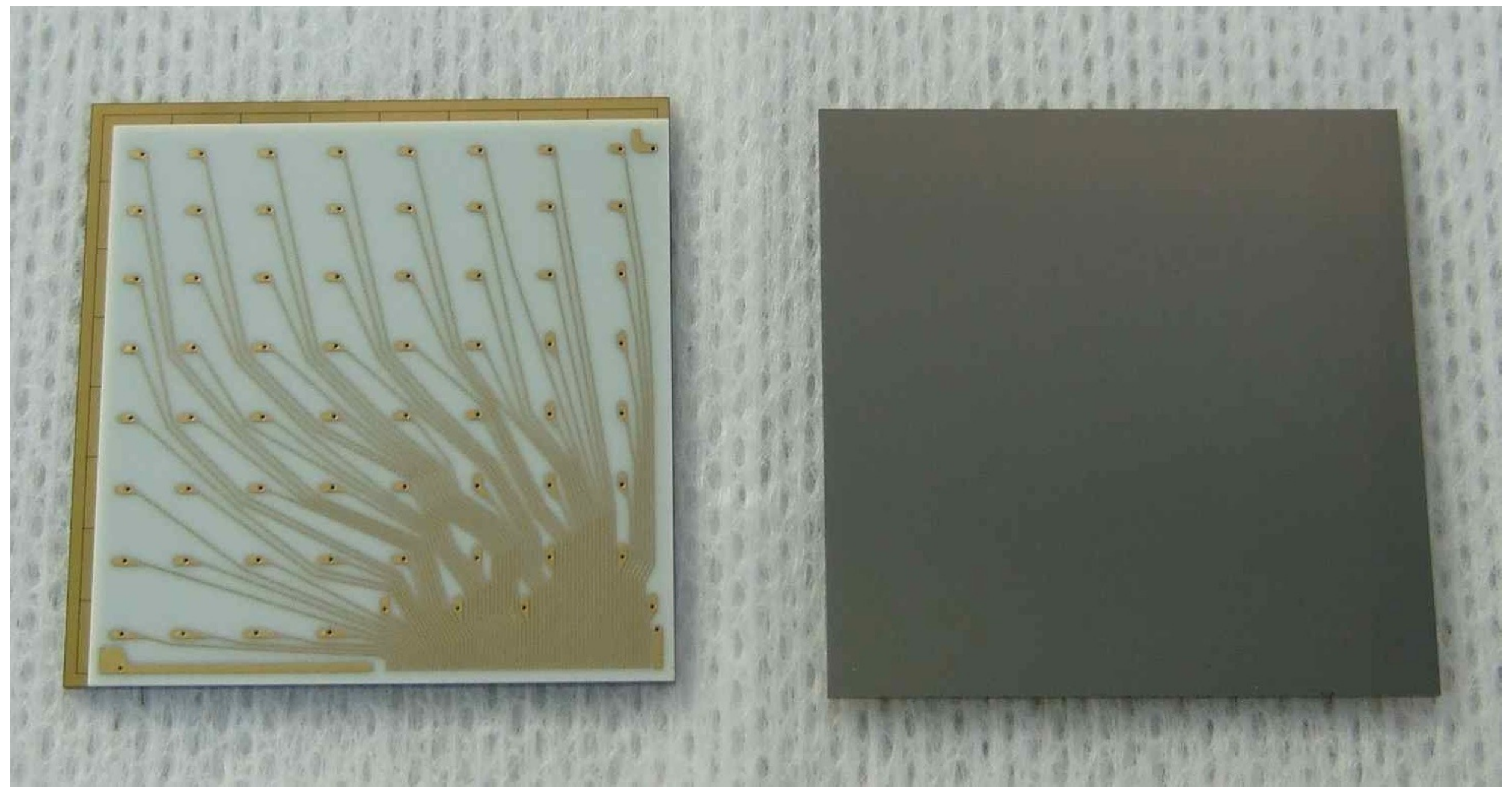} 
   \caption{The CdTe pixel sensor devices bonded with the ceramic fanout board. The left half shows the front with the ceramic
   fanout board for the pixels and the right half shows the common anode electrode utilizing Indium.}
   \label{fig:cdte_sensors}
\end{figure}

Unlike Si sensors, CdTe sensors cannot have integrated readout electrodes above pixel electrodes on the device.
In addition, it is difficult to perform wire-bonding on the electrodes 
of the CdTe sensor. In order to address these issues, we employ 
a separate fanout board to route signal from each pixel to the 
corner of the sensor where ASICs are placed.  The fanout 
board is made of 0.3~mm thick ceramic (Al$_2$O$_3$) substrate 
that allows fine pitch between electrodes to match the input 
pitch of the ASIC (91~$\mu$m).  The CdTe sensor and the fanout board are bonded by using a conductive epoxy 
as shown in Figure~\ref{fig:cdte_sensors}. ASIC and the fanout board are connected by wire bonding.
The readout electronics are DC-coupled to the CdTe sensor.

\begin{figure*}[htbp] 
   \centering
   \begin{tabular}{ll}
   (a) & (b)\\
   \includegraphics[width=0.45\textwidth]{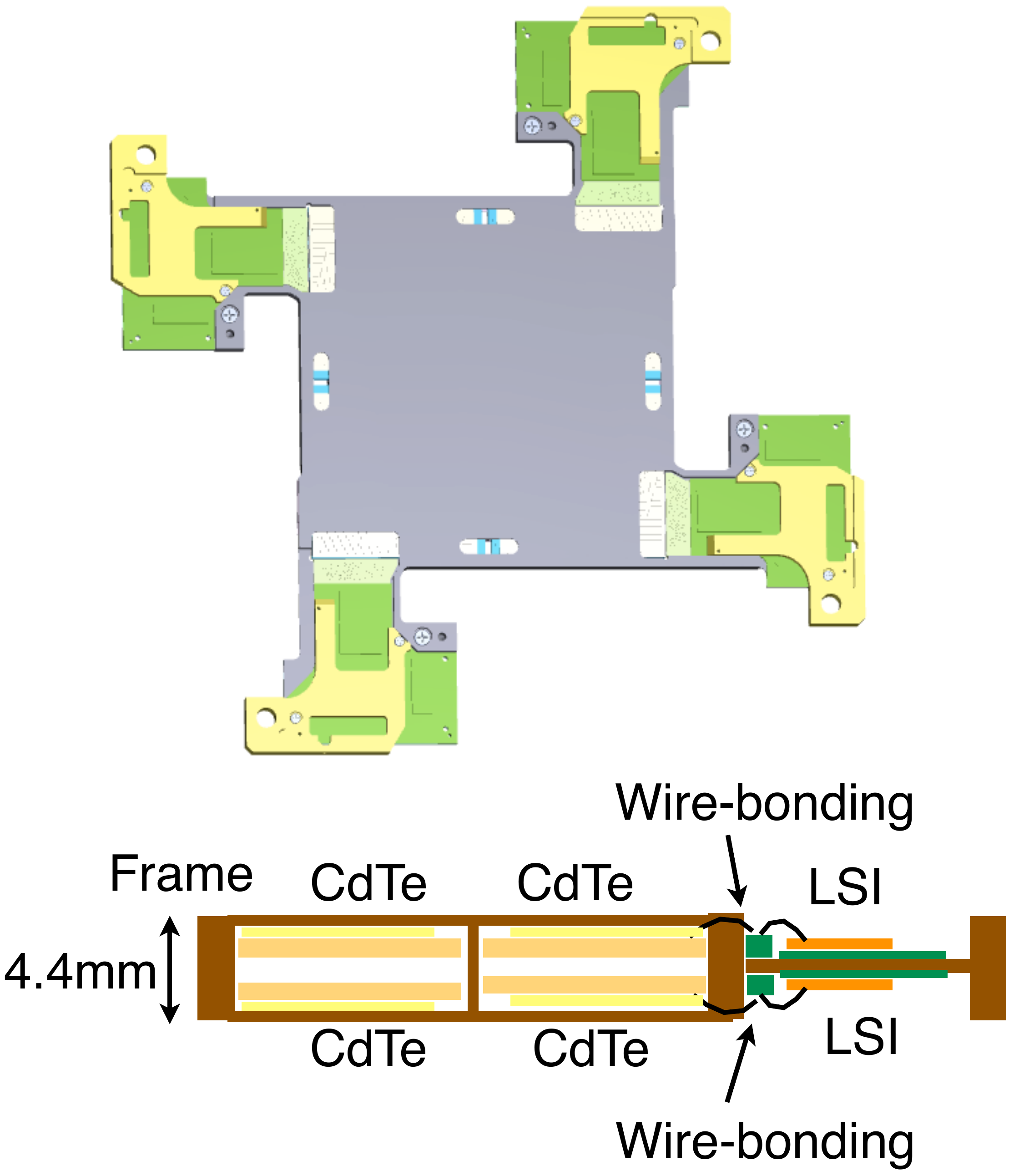} &
   \includegraphics[width=0.45\textwidth]{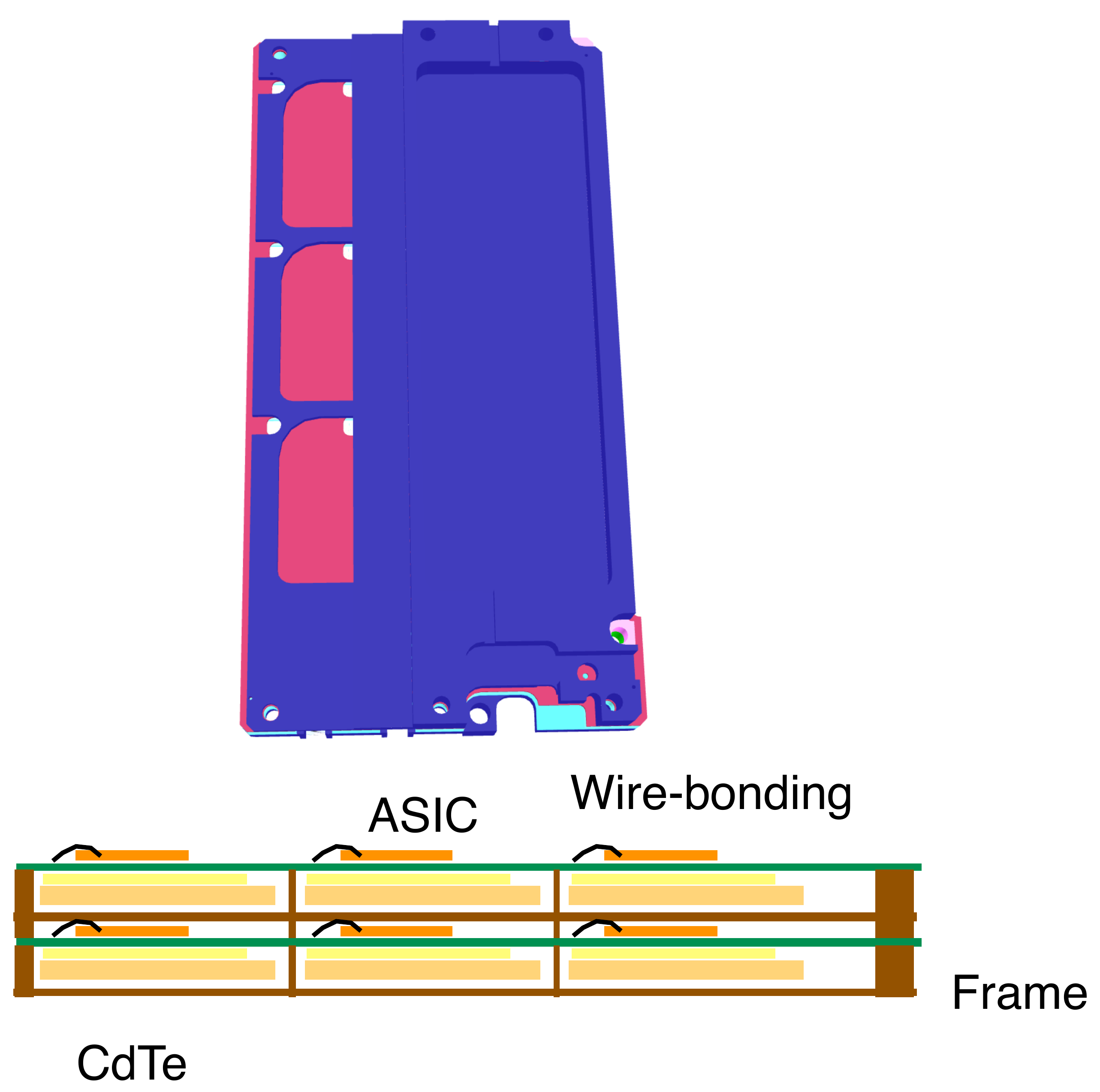}\\
   \end{tabular}   
   \caption{(a) 3D model and conceptual illustration of the stacked CdTe sensor tray module. (b) 3D model and conceptual illustration of the side CdTe sensor tray module.}
   \label{fig:cdte_tray_module}
\end{figure*}

Figure~\ref{fig:cdte_tray_module} (a) shows the structure of the stack CdTe sensor tray module.
This CdTe sensor tray module has the same outer shape as the Si sensor tray module. The stacked CdTe sensor tray module
consists of two $2\times2$ tiled CdTe sensor arrays and eight FECs. In one Compton camera, 
four stacked CdTe sensor tray modules are placed underneath the Si sensor tray modules.

The side CdTe sensors surrounding the Si and CdTe stack play important roles in the SGD Compton camera.
Since the Si part is tall, the side sensors are essential for covering the large solid angle of the Si sensors.
Relatively low energy gamma-ray photons, such as 100--200~keV, are the main target of the SGD and these photons 
are the most likely to be Compton scattered with a large scattering angle. Moreover, these Compton scatterings
with a large scattering angle provide the gamma-ray polarization information\cite{Lei1997309,Takeda10}. 
Figure~\ref{fig:cdte_tray_module} (b) shows the structure of the side CdTe sensor tray module.
This module consists of two layers. CdTe sensors are tiled in an $2\times3$ array for each layer.

\begin{figure}[bthp] 
   \centering
   \includegraphics[width=0.48\textwidth]{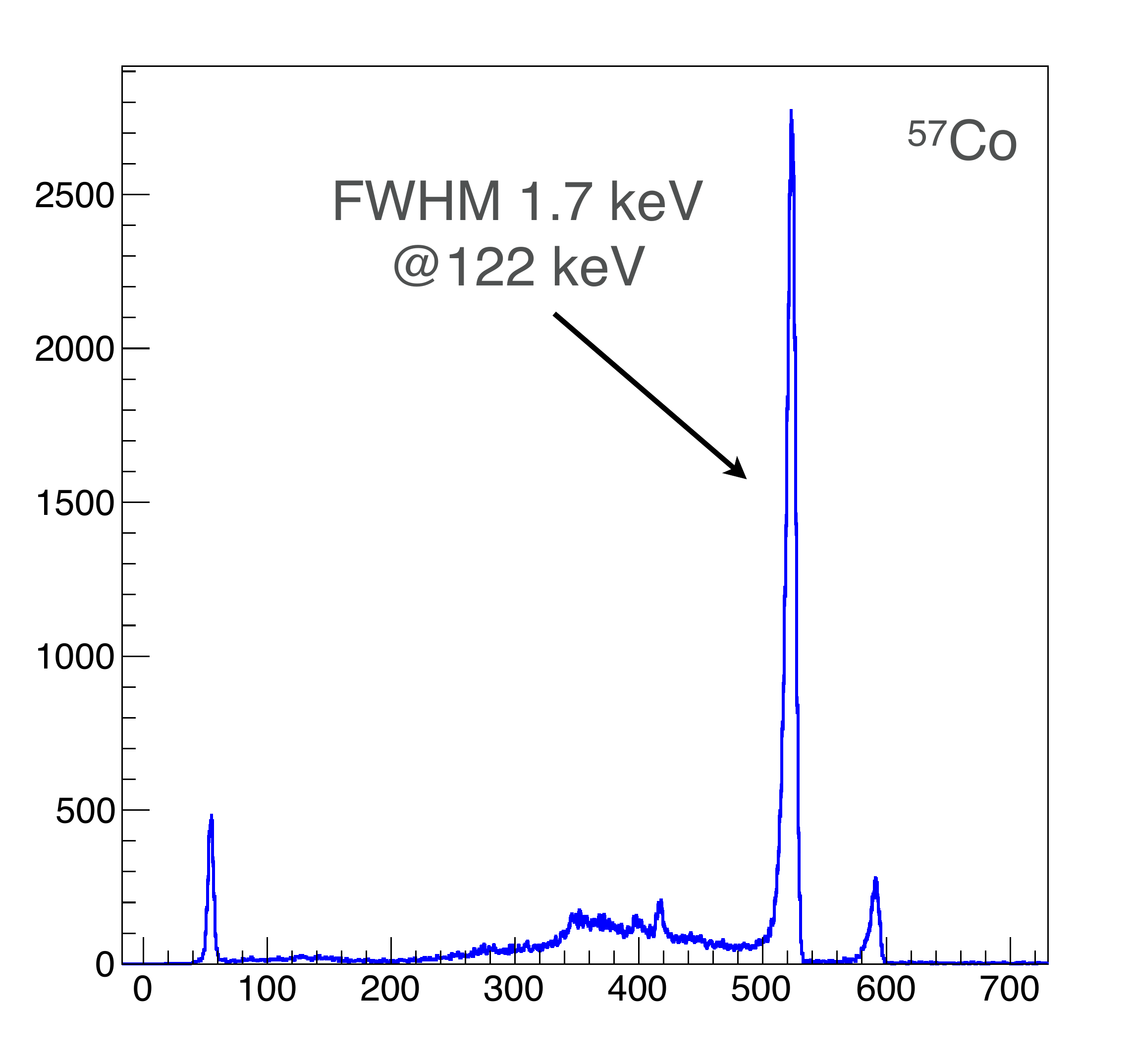} 
   \caption{$^{57}$Co spectrum obtained with the CdTe sensor tray. This spectrum is obtained with one of the best pixels. The FWHM energy resolution is 1.7~keV at 122~keV. The operating temperature is $-$10$^{o}$C and the applied bias voltage is 1000~V. }
   \label{fig:cdtesensors_performance}
\end{figure}

The spectral performance obtained with the CdTe sensor tray is shown in Figure~\ref{fig:cdtesensors_performance}.
This is a $^{57}$Co spectrum obtained with one of the best pixels. The FWHM energy resolution is 1.7~keV at 122~keV. 
The operating temperature is $-$10$^{o}$C and the applied bias voltage is 1000~V.

\subsection{Readout System}
\label{subsec:readoutsystem}
\begin{figure*}[htbp] 
   \centering
   \includegraphics[width=0.82\textwidth]{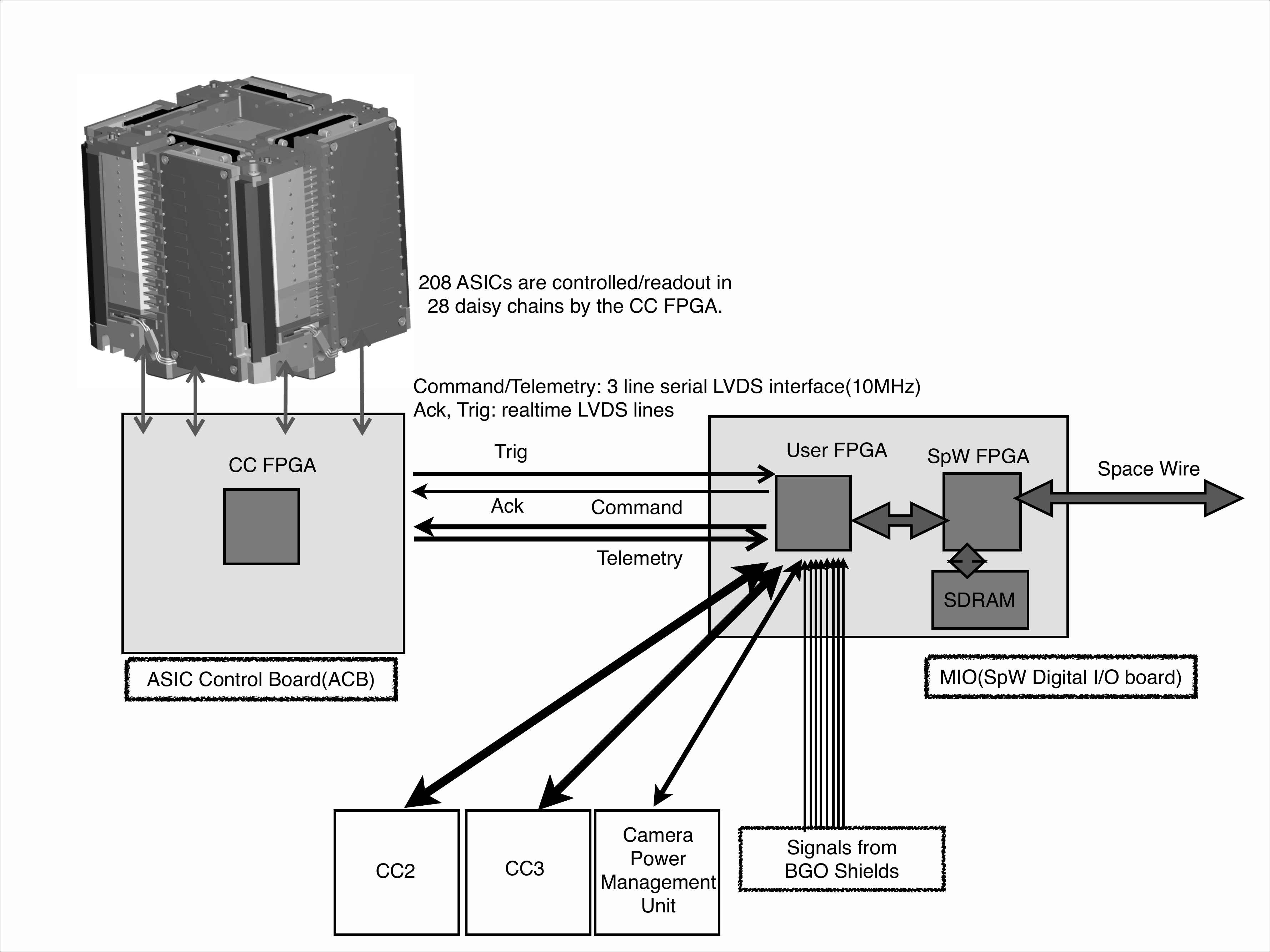} 
   \caption{Readout system of the SGD Compton cameras.}
   \label{fig:cc_readout_system}
\end{figure*}

The readout system of the SGD Compton camera consists of the FECs, ADBs, ACB and MIO(Mission I/O) board as shown in
Figure~\ref{fig:cc_readout_system}.

The front-end electronics of the Compton camera consists of four groups of 42 FECs and an ADB, and an ACB. 
Two FECs are connected back-to-back at the corner of each Si sensor tray module and each stack CdTe sensor tray module, 
and are read out in daisy chain. 
FECs for the side CdTe sensor tray modules have six ASICs that are daisy-chained on each board.
Forty FECs from the Si and stack CdTe trays and two FECs from the side CdTe trays are connected to an ADB, 
which is located on the side of the Compton camera.
Eight FECs (eight ASICs) are daisy-chained for Si and stack CdTe trays, resulting in seven groups of ASICs for each side, 
five for Si and stack CdTe trays (eight ASICs each) and two for side CdTe trays (six ASICs each).
Only digital communication is required between ADBs and FECs and all digital signals are differential to 
minimize the electromagnetic interference.
Digital signals that are not used frequently are single ended between the ADB and the ACB due to 
constraints on the cable pin count.
The ADB detects excess current of each ASIC group in order to protect ASICs from latch-ups due to 
highly ironizing radiation or other causes.
We can recover ASICs from latch-ups by cycling the power supply.
ASICs are controlled by an FPGA on the ACB (one board per Compton camera), and the ACB FPGAs are controlled by the user
FPGA on the MIO board. Communication between the Compton camera and MIO is handled via 3-line (CLK, DATA, STRB) 
serial protocol on LVDS physical layer. We have two additional real-time LVDS lines dedicated for trigger and trigger acknowledgement signals.

\begin{figure*}[hbtp] 
   \centering
   \includegraphics[width=0.95\textwidth]{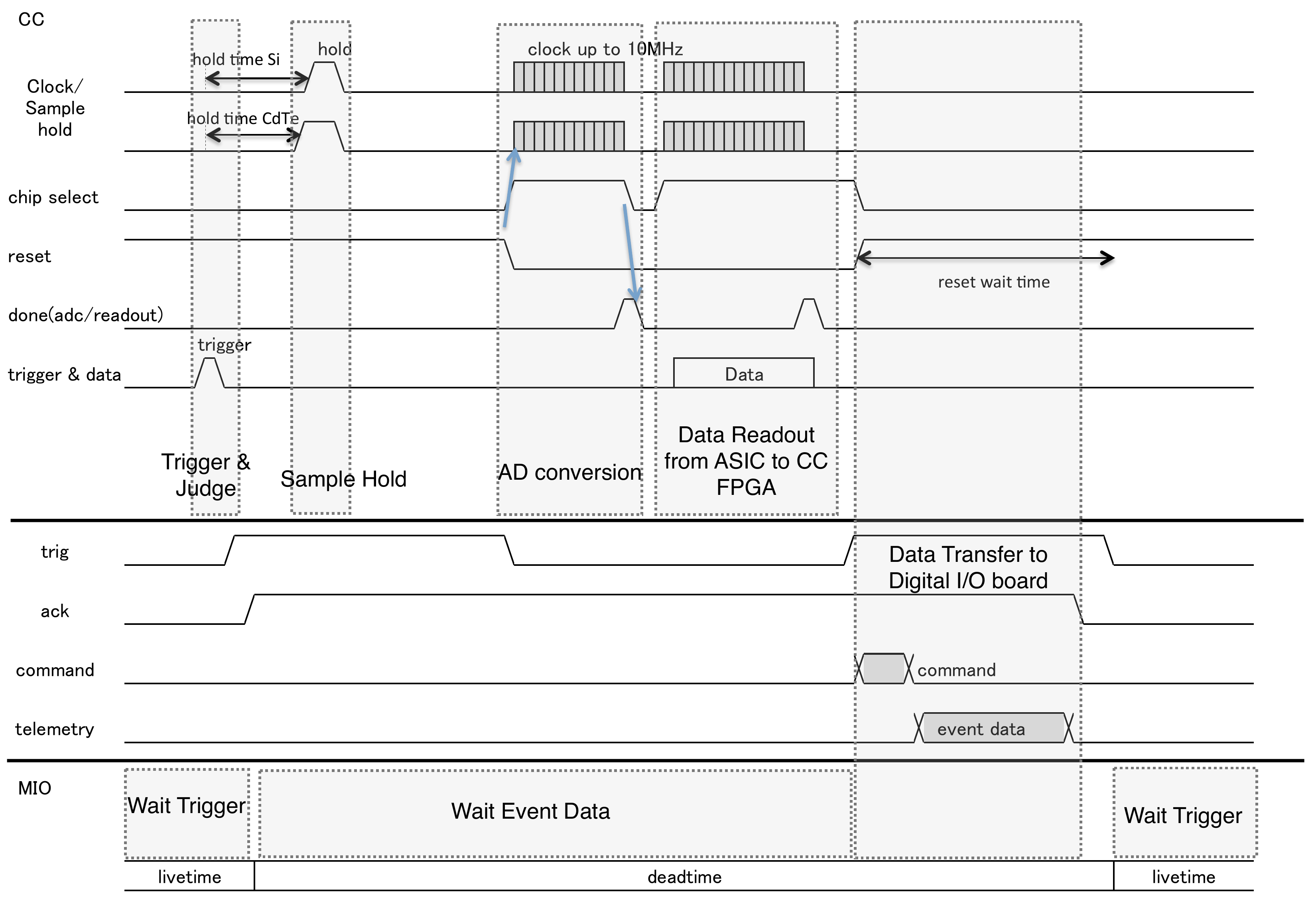} 
   \caption{Timing chart representing the readout sequence of the SGD Compton camera.}
   \label{fig:cc_readout_timingchart}
\end{figure*}

Figure~\ref{fig:cc_readout_timingchart} shows a timing chart representing the readout sequence of the SGD Compton camera.
Once the ACB FPGA has received a trigger signal from an ASIC, the ACB FPGA sends a trigger to the MIO and prepares the sample hold signal for ASICs. After receiving the acknowledgement signal from the MIO, the ACB FPGA holds ASICs' pulse height signals with proper delays upon reception of triggers from ASICs, and controls analog-to-digital conversion on ASICs and data transfer from ASICs. 
The ACB FPGA has an internal memory for one event data, and saves the data transferred from ASICs. 
The MIO FPGA receives the signal from the ACB that the event data has been prepared, and then MIO FPGA read out the data by using the command and telemetry serial lines. In the MIO, the data is formatted and stored in the SDRAM, which can buffer the event data over periods of several seconds. Finally, the buffered event data in the SDRAM are pulled by a data acquisition computer via the Space Wire network.

The dead-time associated with the event data acquisition is not short (100~$\mu$s -- several 100~$\mu$s).
There are two functions for more effective event data acquisitions. One is "Trigger pattern judging". The trigger patterns of 28 daisy chains are categorized into 16 groups. For each group, we can select whether the trigger is sent to the MIO or not. By using this function, we can ignore the charged particle events resulting in no related dead-time.
The other function is FastBGO cancel. FastBGO is the BGO scintillator signal processed with a relatively short time interval ($\sim$~5~$\mu$s) and corresponds to the relatively large energy deposit in one of the BGO shields. With this function, we can cancel the event data acquisition before AD conversion by FastBGO signals, and can reduce the dead-time.  

It is very important for astronomical observations to measure the dead-time accurately. 
To estimate the dead-time, the SGD Compton camera has a ``pseudo trigger'' function.
This function has been successfully introduced in the HXD onboard the Suzaku satellite\cite{Kokubun06_HXD}.
In the SGD Compton camera system, the pseudo triggers are generated by the ACB FPGAs internally
and are then processed in the same manner as usual triggers.
Since the pseudo events are discarded 
if the pseudo trigger is generated 
while a ``real event'' is inhibiting other triggers, 
the dead-time fraction can be estimated by counting 
a number of pseudo events, output to the telemetry,
and comparing with the expected counts during the same exposure.
Although periodic pseudo triggers were used in the HXD system, 
the random pseudo trigger function is developed for the SGD system.
A random pseudo trigger function is based on the pseudorandom numbers calculated in the FPGA and 
the expected count rate can be set up to 422~Hz.
Random pseudo trigger function enable us to estimate the dead-time due to the 
telemetry saturation in the back-end network, which can occur periodically.

\section{Performance evaluation with the final prototype}
\label{sec:performance}

\subsection{Final prototype}
\label{subsec:finalprototype}
\begin{figure}[bthp] 
   \centering
   \includegraphics[width=0.45\textwidth]{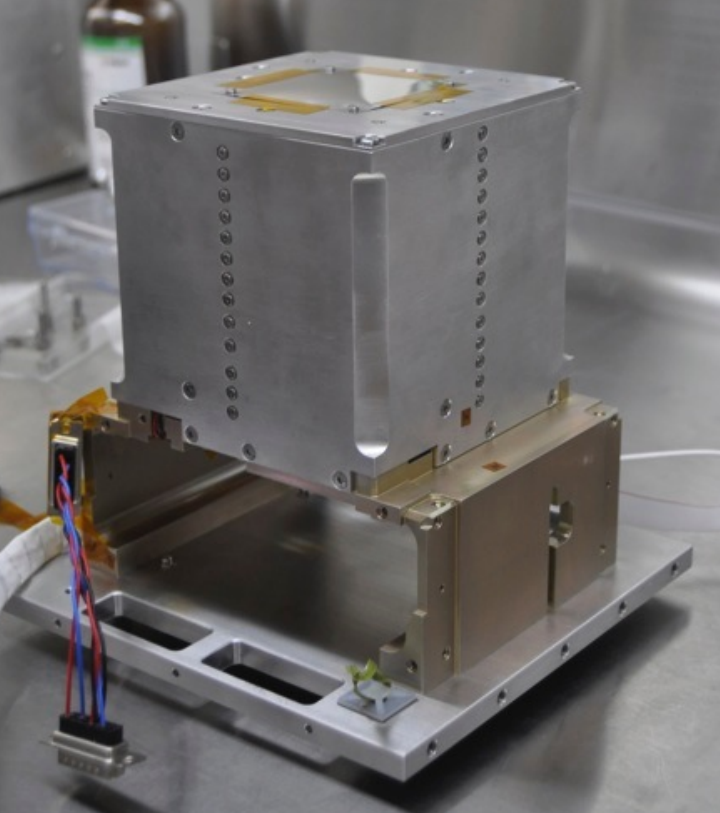} 
   \caption{Picture of the final prototype of the SGD Compton camera.}
   \label{fig:final_prototype}
\end{figure}

Prior to the production of the flight model, we have fabricated the final prototype of the SGD Compton camera.
Figure~\ref{fig:final_prototype} shows a picture of the prototype. The prototype has all the  
components installed in the same manner as the flight model except for some space-qualified electronic parts. 

\subsection{Spectral Performance}
\label{subsec:spectral_performance}

In order to evaluate the spectral performance of the prototype, we performed gamma-ray measurement tests
using various radio isotopes. In the measurement tests, the prototype was operated in the temperature of about $-$10$^{\circ}$C.
The prototype camera was cooled in a portable freezer and was irradiated with gamma rays by radio isotopes outside of the freezer.

\begin{figure*}[htbp] 
   \centering
   \includegraphics[width=0.98\textwidth]{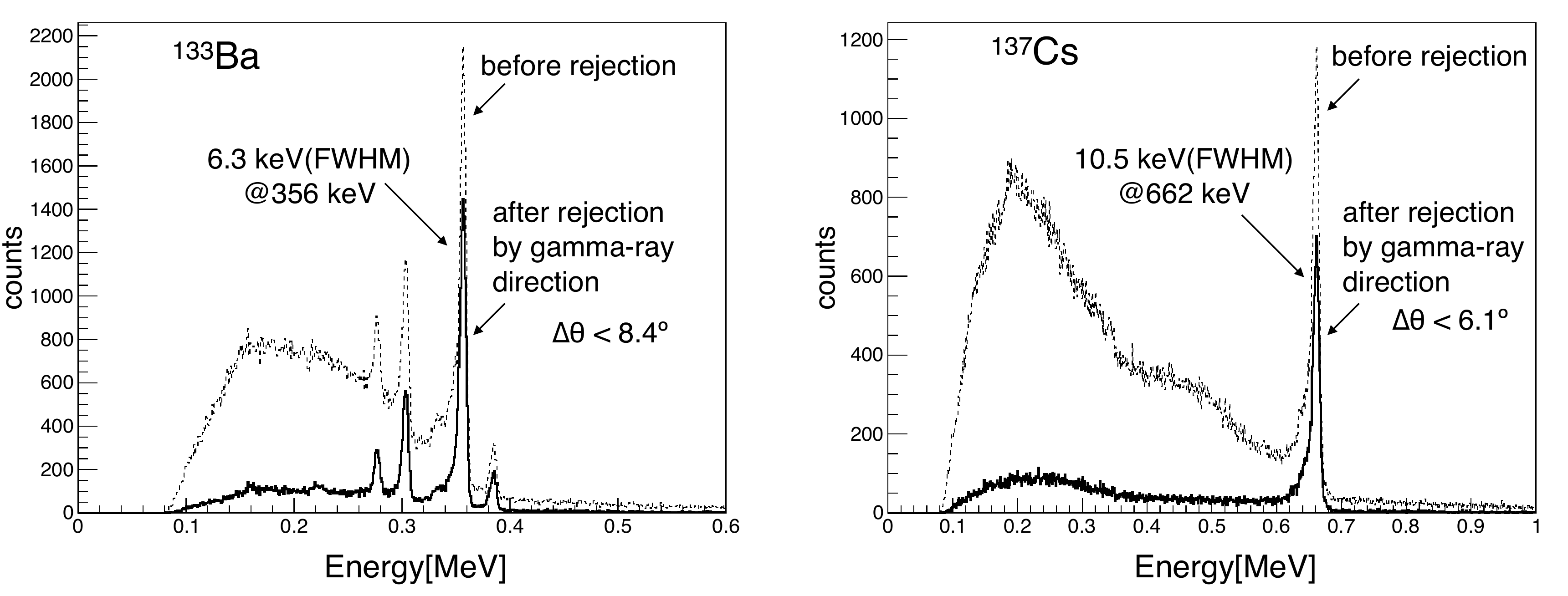} 
   \caption{The gamma-ray spectra of $^{133}$Ba (left) and $^{137}$Cs (right) obtained with the final prototype
   of the SGD Compton camera. The event data applicable to the Compton calculation are used for the dotted line spectra.
   The FWHM energy resolutions at 356~keV and 662~keV are 6.3~keV and 10.5~keV, respectively.
   The solid lines show the spectra obtained after the rejection by gamma-ray incident direction, which are calculated 
   by the information of Compton scattering. The selection criteria for $^{133}$Ba and $^{137}$Cs are 
   $\Delta \theta < 8.4^{\circ}$ and $\Delta \theta < 6.1^{\circ}$, respectively ($\Delta \theta$ shows the angular resolution measure as 
   $\Delta \theta = | \theta_\mathrm{energy} - \theta_\mathrm{geom}|$.) }
   \label{fig:cc_spectral_performance}
\end{figure*}

$^{133}$Ba and $^{137}$Cs gamma-ray spectra obtained with the prototype are 
shown in Figure~\ref{fig:cc_spectral_performance}.
The dotted line spectra are made from the event data 
that can be used to reconstruct the incident gamma-ray information using Compton kinematics.
The obtained FWHM energy resolution is 6.3~keV and 10.5~keV for gamma rays of 356~keV and 662~keV, respectively.
These spectral characteristics satisfy the requirement for the SGD Compton camera, better than 2\%(FWHM).

By using the information gained from Compton scattering, we can obtain a constraint of the incident angle and
can select event data from gamma rays that enter the camera directly from the radio isotope. 
The solid line spectra in Figure~\ref{fig:cc_spectral_performance} are made from
the event data after incident direction selection.
This selection is performed by using the angular resolution measure (ARM) as 
$\Delta \theta = | \theta_\mathrm{energy} - \theta_\mathrm{geom}|$. Here, $\theta_\mathrm{energy}$ is the Compton-scattering
angle calculated from the detected energy information, and $\theta_\mathrm{geom}$ is that determined from the detected position 
information. The selection criteria for $^{133}$Ba and $^{137}$Cs are $\Delta \theta < 8.4^{\circ}$ and $\Delta \theta < 6.1^{\circ}$,
respectively.
The spectral components besides emission lines are compressed with the selection, because
these components are mainly generated by the gamma rays scattered by the materials around the camera 
and the radio isotope.
These results provide partial confirmation about background rejection capabilities of SGD and SGD Compton camera. 

The $\sim$~100 hours long measurement tests have been performed. During the tests, 
we have confirmed that there is no significant variation in performance.
Moreover, a thermal vacuum test simulating the environment in space has been performed. 
The final prototype Compton camera was put into a thermal vacuum chamber with a vacuum pressure of $\sim$~2~$\times$~10$^{-3}$~Pa and
a temperature of $-$15$^{\circ}$C -- $-$25$^{\circ}$C, and, the stable operation in this environment was confirmed.

\subsection{Effective Area}
\label{subsec:effarea}

The response functions for gamma-ray photons including the effective area are built with a Monte Carlo simulator.
We have constructed the Monte Carlo simulator and have tuned it by adopting the measurement results 
from the final prototype of SGD.
The details of the Monte Carlo simulators for early phase prototypes were described in \cite{Takeda09,Odaka07,Takeda10,Odaka10_MC}, and 
the details of the current SGD Monte Carlo simulator and event reconstruction algorithm will be described 
in a separate publication.

\begin{figure}[bthp] 
   \centering
   \includegraphics[width=0.48\textwidth]{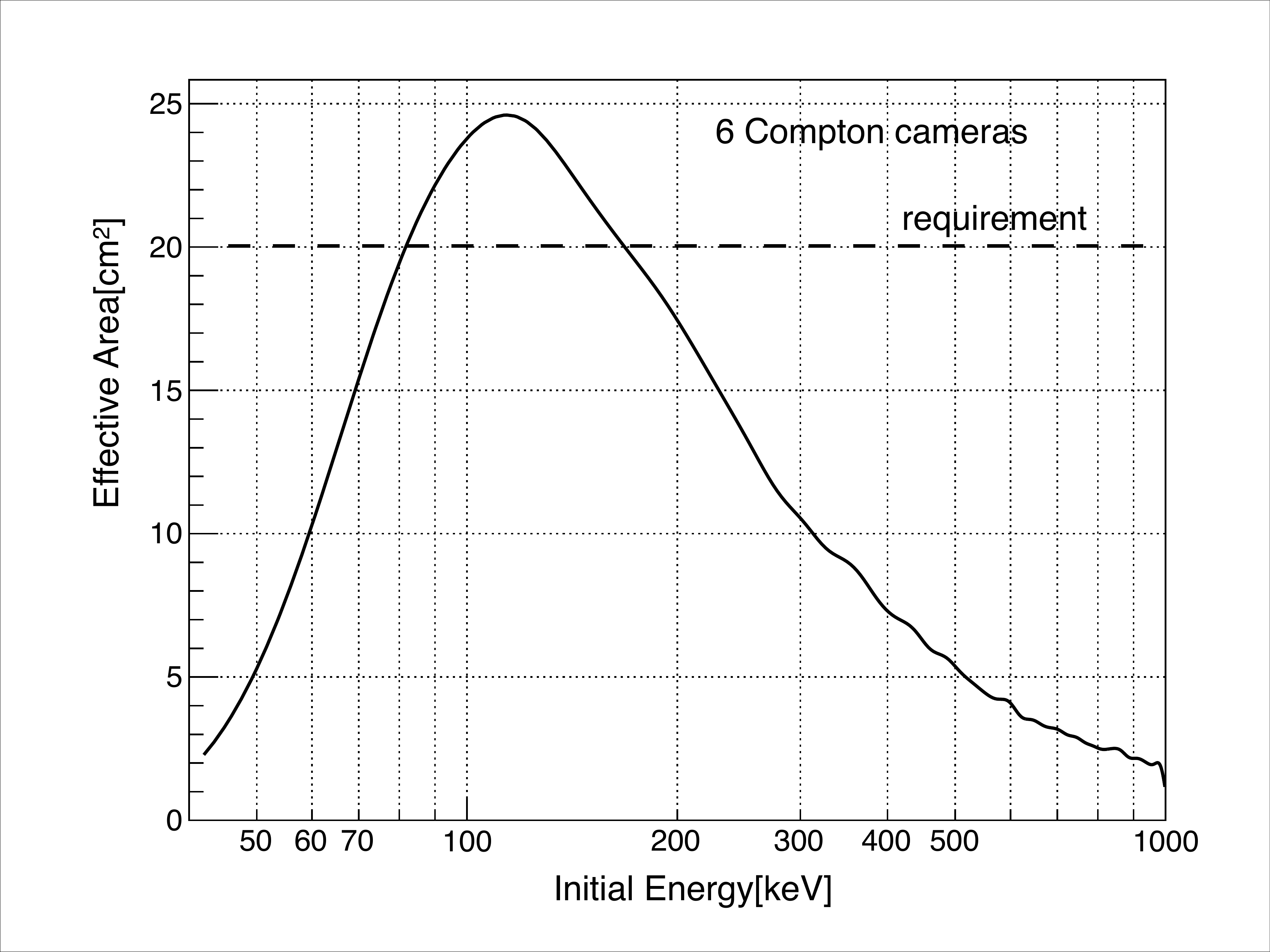} 
   \caption{Plot of the effective area for six SGD Compton cameras onboard ASTRO-H. The effective area is derived from the Monte Carlo simulator adopting the measurement results with the prototype.}
   \label{fig:cc_effarea}
\end{figure}

The effective area based on the Monte Carlo simulator is shown in Figure~\ref{fig:cc_effarea}.
The effective area is defined as the product of the detector geometrical area and the detection efficiency, 
which is derived from the Monte Carlo simulation. 
The combined effective area for the six Compton cameras is estimated to be larger than 20~cm$^2$ at 100~keV, and this value 
satisfies the instrument-level requirements for the SGD Compton camera.
The efficiency reaches about 15\% and 3\% for 100~keV and 511~keV gamma rays, respectively.

\subsection{Intrinsic background}
\label{subsec:intrinsicbg}

Low background is the most important factor for sensitive observations with the SGD.
Therefore, we performed background measurements with the prototype
in order to evaluate the intrinsic background generated by radio isotopes
contained in the SGD Compton camera.
The measurements were held in a lead cave, which works as a passive shield.
This lead cave consists of 5~cm thick lead with 1--5~cm of thick copper or brass layer inside the lead shield.
The prototype Compton camera with the Pb cave was cooled to $-$10$^{\circ}$C in a thermostatic chamber.

\begin{figure}[bthp] 
   \centering
   \includegraphics[width=0.48\textwidth]{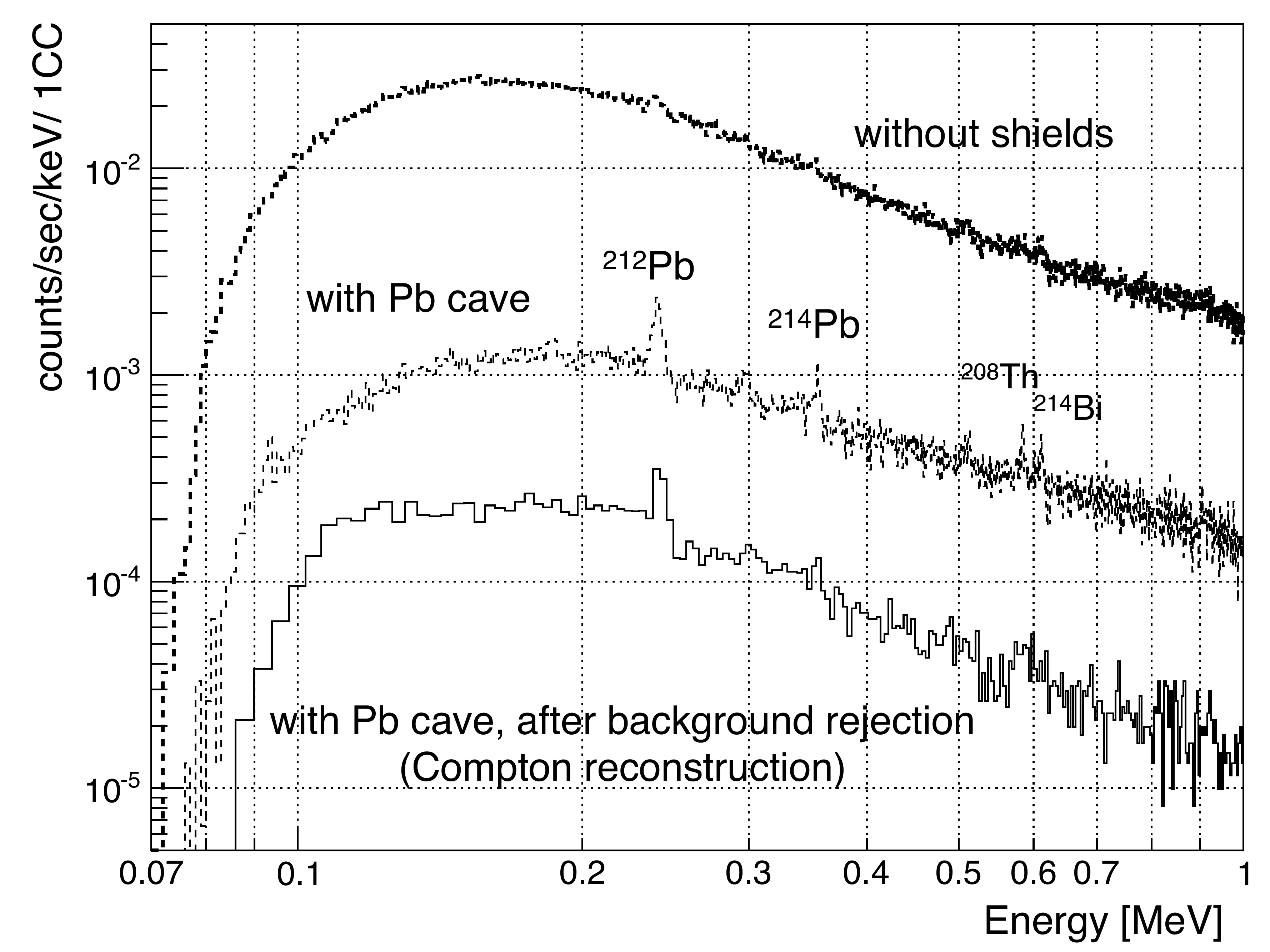} 
   \caption{Background spectra obtained with the prototype SGD Compton camera. The room background spectrum obtained without any shields, the spectrum with the Pb cave, and the spectrum after the Compton reconstruction are plotted.}
   \label{fig:cc_intrinsicbg}
\end{figure}

The obtained spectra are shown in Figure~\ref{fig:cc_intrinsicbg} together with the background
spectrum obtained without any shields. 
Although some emission lines from the thorium and uranium decay series can be seen,
the background level per effective area is as low as several~$\times$~10$^{-5}$ counts/sec/keV/cm$^2$ after the Compton reconstruction.
We have confirmed that there is no strong contaminating background and will perform further measurements
using BGO active shields.

\section{Summary}

We have finalized the design of the SGD Compton camera.
The Compton camera has an overall size of 12~cm~$\times$~12~cm~$\times$~12~cm, 
consisting of 32 layers of Si pixel sensors and 8 layers of CdTe pixel sensors surrounded 
by 2 layers of CdTe pixel sensors.
The pixel pitch of the Si and CdTe sensors is 3.2~mm, and the signals from 13312 pixels in total
are processed by 208 readout ASICs with low noise and low power consumption.
In order to evaluate the performance of the Compton camera, we have fabricated a final prototype,
which has the same design as the flight model. 
The final prototype has been evaluated in the laboratory and the spectral performance, effective area
and low background have been confirmed.

\section*{Acknowledgements}
This work was supported by JST-SENTAN program and JSPS KAKENHI Grant Number 21684015, 24105007, 24244014, 24244021.
The authors would like to thank Ms. Kirsten Bonson and Dr. Poshak Gandhi for their critical reading of the manuscript.


\begin{thebibliography}{10}
\expandafter\ifx\csname url\endcsname\relax
  \def\url#1{\texttt{#1}}\fi
\expandafter\ifx\csname urlprefix\endcsname\relax\def\urlprefix{URL }\fi
\expandafter\ifx\csname href\endcsname\relax
  \def\href#1#2{#2} \def\path#1{#1}\fi

\bibitem{NeXT08}
T.~Takahashi, et~al., The {NeXT} {X}-ray mission, new exploration {X}-ray
  telescope, Proc. SPIE 7011 (2008) 14T.

\bibitem{Takahashi10}
T.~Takahashi, K.~Mitsuda, R.~L. Kelley, et~al., The {ASTRO-H} mission, Proc.
  SPIE 7732 (2010) 77320Z--77320Z--18.

\bibitem{Takahashi12}
T.~Takahashi, K.~Mitsuda, R.~L. Kelley, et~al., The {ASTRO-H} x-ray
  observatory, Proc. SPIE 8443 (2012) 844312.

\bibitem{Takahashi04NAR}
T.~Takahashi, K.~Makishima, Y.~Fukazawa, M.~Kokubun, K.~Nakazawa, M.~Nomachi,
  H.~Tajima, M.~Tashiro, Y.~Terada, Hard X-ray and $\gamma$-ray detectors for
  the next mission, New Astronomy Reviews 48 (2004) 269--273.

\bibitem{Takahashi02-NeXT}
T.~Takahashi, T.~Kamae, K.~Makishima, Future hard {X}-ray and gamma-ray
  observations, Proc. SPIE 251 (2002) 210--213.

\bibitem{Takahashi02}
T.~Takahashi, K.~Nakazawa, T.~Kamae, H.~Tajima, Y.~Fukazawa, M.~Nomachi,
  M.~Kokubun, High resolution {CdTe} detectors for the next generation
  multi-{Compton} gamma-ray telescope, in: J.~E. Truemper, H.~D. Tananbaum
  (Eds.), X-ray and Gamma-ray Telescopes and Instruments for Astronomy, {SPIE},
  Vol. 4851, 2002, pp. 1228--1235.

\bibitem{Takahashi04-SGD}
T.~Takahashi, A.~Awaki, T.~Dotani, Y.~Fukazawa, K.~Hayashida, T.~Kamae,
  J.~Kataoka, N.~Kawai, et~al., Wide-band {X}-ray imager ({WXI}) and soft
  gamma-ray detector ({SGD}) for the {NeXT} mission, Proc. SPIE 5488 (2004)
  549--560.

\bibitem{HXI08}
M.~Kokubun, et~al., Hard {X}-ray imager {HXI} for the {NeXT} mission, Proc.
  SPIE 7011 (2008) 21K.

\bibitem{Kokubun10}
M.~Kokubun, S.~Watanabe, K.~Nakazawa, H.~Tajima, Y.~Fukazawa, T.~Takahashi,
  J.~Kataoka, T.~Kamae, H.~Katagiri, G.~Madejski, K.~Makishima, T.~Mizuno,
  M.~Ohno, R.~Sato, H.~Takahashi, T.~Tanaka, M.~Tashiro, Y.~Terada, K.~Yamaoka,
  the HXI/SGD~team, Hard X-ray and gamma-ray detector for astro-h based on Si
  and CdTe imaging sensors, Nucl. Instrum. Methods A 623 (2010) 425--427.

\bibitem{Kokubun10spie}
M.~Kokubun, et~al., Hard X-ray imager for the {ASTRO-H} mission, Proc. SPIE
  7732 (2010) 773215--773215--13.

\bibitem{Kokubun12spie}
M.~Kokubun, et~al., The hard X-ray imager (HXI) for the {ASTRO-H} mission,
  Proc. SPIE 8443 (2012) 844325.

\bibitem{Tajima05}
H.~Tajima, T.~Kamae, G.~Madejski, T.~Mitani, K.~Nakazawa, T.~Tanaka,
  T.~Takahashi, S.~Watanabe, et~al., Design and performance of the {Soft
  Gamma-ray Detector} for the {NeXT} mission, IEEE Trans. Nucl. Sci. 53 (2005)
  2749--2757.

\bibitem{Tajima10}
H.~Tajima, et~al., Soft gamma-ray detector for the {ASTRO-H} mission, Proc.
  SPIE 7732 (2010) 73216--773216--17.

\bibitem{Watanabe12spie}
S.~Watanabe, et~al., Soft gamma-ray detector for the {ASTRO-H} mission, Proc.
  SPIE 8443 (2012) 844326.

\bibitem{Tajima02}
H.~Tajima, T.~Kamae, S.~Uno, T.~Nakamoto, Y.~Fukazawa, T.~Mitani, T.~Takahashi,
  K.~Nakazawa, Y.~Okada, M.~Nomachi, Low noise double-sided silicon strip
  detector for multiple-{Compton} gamma-{ray} telescope, in: J.~E. Truemper,
  H.~D. Tananbaum (Eds.), X-ray and Gamma-ray Telescopes and Instruments for
  Astronomy, {SPIE}, Vol. 4851, 2002, pp. 875--884.

\bibitem{Tajima03}
H.~Tajima, Gamma-ray polarimetry, Nucl. Instrum. Methods A 511 (2003) 287--290.

\bibitem{Fukazawa05_si}
Y.~Fukazawa, T.~Nakamoto, N.~Sawamoto, S.~Uno, T.~Ohsugi, H.~Tajima,
  T.~Takahashi, T.~Mitani, T.~Tanaka, K.~Nakazawa, Development of low-noise
  double-sided silicon strip detector for cosmic soft gamma-ray Compton camera,
  Nucl. Instr. and Meth. A 541 (2005) 342--349.

\bibitem{Takeda07}
S.~Takeda, S.~Watanabe, T.~Tanaka, K.~Nakazawa, T.~Takahashi, Y.~Fukazawa,
  H.~Yasuda, H.~Tajima, Y.~Kuroda, M.~Onishi, K.~Genba, Development of
  double-sided silicon strip detectors (DSSD) for a Compton telescope, Nucl.
  Instrum. Methods A 579 (2007) 859--865.

\bibitem{Takahashi01b}
T.~Takahashi, S.~Watanabe, Recent progress in {CdTe} and {CdZnTe} detectors,
  IEEE Trans. Nucl. Sci. 48 (2001) 950--959.

\bibitem{Watanabe05}
S.~Watanabe, T.~Tanaka, K.~Nakazawa, T.~Mitani, K.~Oonuki, T.~Takahashi,
  T.~Takashima, H.~Tajima, Y.~Fukazawa, M.~Nomachi, S.~Kubo, M.~Onishi,
  Y.~Kuroda, A {Si/CdTe} semiconductor {Compton} camera, IEEE Trans. Nucl. Sci.
  52 (2005) 2045--2051.

\bibitem{Watanabe09}
S.~Watanabe, S.~Ishikawa, H.~Aono, S.~Takeda, H.~Odaka, M.~Kokubun,
  T.~Takahashi, K.~Nakazawa, H.~Tajima, M.~Onishi, Y.~Kuroda, High energy
  resolution hard {X-ray} and gamma-ray imagers using {CdTe} diode devices,
  IEEE Trans. Nucl. Sci. 56 (2009) 777--782.

\bibitem{Takeda09}
S.~Takeda, H.~Aono, S.~Okuyama, S.~n.~Ishikawa, H.~Odaka, S.~Watanabe,
  M.~Kokubun, T.~Takahashi, K.~Nakazawa, H.~Tajima, N.~Kawachi, Experimental
  results of the gamma-ray imaging capability with a {Si/CdTe} semiconductor
  {Compton} camera, IEEE Trans. Nucl. Sci. 56 (2009) 783--790.

\bibitem{Takeda12}
S.~Takeda, H.~Odaka, S.~Ishikawa, S.~Watanabe, H.~Aono, T.~Takahashi,
  Y.~Kanayama, M.~Hiromura, S.~Enomoto, Demonstration of in-vivo multi-probe
  tracker based on a {Si/CdTe} semiconductor {Compton} camera, IEEE Trans.
  Nucl. Sci. 59 (2012) 70--76.

\bibitem{CdTebook1}
T.~Takahashi, S.~Watanabe, S.~Ishikawa, High-resolution CdTe detectors and
  application to gamma-ray imaging, Semiconductor Radiation Detection Systems
  CRC Press (2010) Chapter 8.

\bibitem{CdTebook2}
T.~Takahashi, S.~Watanabe, S.~Ishikawa, G.~Sato, High-resolution CdTe detectors
  and their application to gamma-ray imaging, Sensor Technologies: Biological
  and Medical Sensors (2011) Chapter 14.

\bibitem{HXD}
T.~Kamae, H.~Ezawa, Y.~Fukazawa, H.~M, E.~Idesawa, N.~Iyomoto, et~al.,
  {Astro-E} hard {X}-ray detector, Proc. SPIE 2806 (1996) 314.

\bibitem{Ribberfors19752067}
R.~Ribberfors, Relationship of the relativistic Compton cross section to the
  momentum distribution of bound electron states, Physical Review B 12~(6)
  (1975) 2067--2074.

\bibitem{Zoglauer20021302}
A.~Zoglauer, G.~Kanbach, Doppler broadening as a lower limit to the angular
  resolution of next generation Compton telescopes, Proceedings of SPIE - The
  International Society for Optical Engineering 4851~(2) (2002) 1302--1309.

\bibitem{VA94}
O.~Toker, S.~Masciocchi, E.~Nyg{\aa}rd, A.~Rudge, P.~Weilhammer, {VIKING}, a
  {CMOS} low noise monolithic 128 channel frontend for {Si}-strip detector
  readout, Nucl. Instrum. Methods A 340 (1994) 572--579.

\bibitem{Tajima04}
H.~Tajima, T.~Nakamoto, T.~Tanaka, S.~Uno, T.~Mitani, E.~do~Couto~e Silva,
  et~al., Performance of a low noise front-end {ASIC} for {Si/CdTe} detectors
  in {Compton} gamma-ray telescope, IEEE Trans. Nucl. Sci. 51 (2004) 842--847.

\bibitem{Barthelmy2005143}
S.~Barthelmy, L.~Barbier, J.~Cummings, E.~Fenimore, N.~Gehrels, D.~Hullinger,
  H.~Krimm, C.~Markwardt, D.~Palmer, A.~Parsons, G.~Sato, M.~Suzuki,
  T.~Takahashi, M.~Tashiro, J.~Tueller, The burst alert telescope (BAT) on the
  Swift MIDEX mission, Space Science Reviews 120~(3-4) (2005) 143--164.

\bibitem{Picozza2007296}
P.~Picozza, et~al., Pamela - a payload for antimatter matter exploration and
  light-nuclei astrophysics, Astroparticle Physics 27~(4) (2007) 296--315.

\bibitem{Tavani2009995}
M.~Tavani, et~al., The AGILE mission, Astronomy and Astrophysics 502~(3) (2009)
  995--1013.

\bibitem{Lei1997309}
F.~Lei, A.~Dean, G.~Hills, Compton polarimetry in gamma-ray astronomy, Space
  Science Reviews 82~(3-4) (1997) 309--388.

\bibitem{Takeda10}
S.~Takeda, H.~Odaka, J.~Katsuta, S.~Ishikawa, S.~Sugimoto, Y.~Koseki,
  S.~Watanabe, G.~Sato, M.~Kokubun, T.~Takahashi, K.~Nakazawa, Y.~Fukazawa,
  H.~Tajima, H.~Toyokawa, Polarimetric performance of {Si/CdTe} semiconductor
  {Compton} camera, Nucl. Instrum. Methods A 622 (2010) 619--627.

\bibitem{Kokubun06_HXD}
M.~Kokubun, et~al., In-orbit performance of the hard X-ray detector onboard
  suzaku, Publications of the Astronomical Society of Japan Vol.59, No.SP1
  (2007) 53--76.

\bibitem{Odaka07}
H.~Odaka, S.~Takeda, S.~Watanabe, S.~Ishikawa, M.~Ushio, T.~Tanaka,
  K.~Nakazawa, T.~Takahashi, H.~Tajima, Y.~Fukazawa, Performance study of
  {Si/CdTe} semiconductor {Compton} telescopes with Monte Carlo simulation,
  Nucl. Instrum. Methods A 579 (2007) 878--885.

\bibitem{Odaka10_MC}
H.~Odaka, S.~Sugimoto, S.~Ishikawa, J.~Katsuta, Y.~Koseki, T.~Fukuyama,
  S.~Saito, R.~Sato, G.~Sato, S.~Watanabe, M.~Kokubun, T.~Takahashi, S.~Takeda,
  Y.~Fukazawa, T.~Tanaka, H.~Tajima, Development of an integrated response
  generator for Si/CdTe semiconductor Compton cameras, Nucl. Instr. and Meth. A
  624 (2010) 303--309.

\end{thebibliography}

\end{document}